\documentclass[preprint]{aastex62}
\usepackage{amsmath,amssymb,CJK,soul}
\usepackage{longtable}
\usepackage{hyperref}
\usepackage{longtable}
\usepackage{threeparttablex} 
\usepackage{lineno}
\usepackage{natbib}

\newcommand{\rw}{{2024 RW$_1$ }}
\newcommand{\rwns}{{2024 RW$_1$}}

\received{--}
\revised{--}
\accepted{--}
\submitjournal{AJ}

\shorttitle{Characterization of primitive Imminent Impactor \rw}
\shortauthors{Ingebretsen et al.}

\begin{document}

\title{Apache Point rapid response characterization of primitive imminent impactor \rw}

\author[0000-0002-7053-5495]{Carl Ingebretsen}
\affil{William H. Miller III Department of Physics and Astronomy, Johns Hopkins University, Baltimore, MD 21218, USA}

\author[0000-0002-4950-6323]{Bryce T. Bolin}
\affiliation{Eureka Scientific, Oakland, CA 94602, USA}

\author[0000-0001-7830-028X]{Robert Jedicke}
\affiliation{Institute for Astronomy, University of Hawai`i at M\={a}noa, Honolulu, HI, 96822, USA}

\author[0000-0002-5396-946X]{Peter Vere\v{s}}
\affiliation{Harvard-Smithsonian Center for Astrophysics, Minor Planet Center, Cambridge, MA 02138, USA}

\author[0000-0002-8382-0447]{Christine H. Chen}
\affiliation{Space Telescope Science Institute, 3700 San Martin Dr., Baltimore, MD 21218, USA}
\affil{William H. Miller III Department of Physics and Astronomy, Johns Hopkins University, Baltimore, MD 21218, USA}


\author[0000-0002-9548-1526]{Carey M. Lisse}
\affiliation{Johns Hopkins University Applied Physics Laboratory, 11100 Johns Hopkins Rd, Laurel, MD 20723, USA}

\author{Russet McMillan}
\affiliation{Apache Point Observatory, Sunspot, NM, 88349 USA}

\author[0009-0007-4277-0360]{Torrie Sutherland}
\affiliation{Apache Point Observatory, Sunspot, NM, 88349 USA}

\author[0000-0002-6377-4869]{Amanda J. Townsend}
\affiliation{Apache Point Observatory, Sunspot, NM, 88349 USA}



\begin{abstract}
Imminent impactors may be detected only a few hours before their impact with Earth, providing a brief opportunity to characterize them before impact. We describe the characterization of imminent impactor \rwns, which was discovered by the Catalina Sky Survey on 2024 September 4 at 05:43 UTC, before it entered the atmosphere near the northern Philippines at 16:39 UTC. We observed \rw with the Astrophysical Research Consortium Telescope Imaging Camera on the Apache Point Astrophysical Research Consortium's 3.5-m telescope on 2024 September 4 10:16 UTC. We obtained g, r, i, and z photometry of \rwns, yielding color indices of g-r = 0.47$\pm$0.04, r-i = 0.13$\pm$0.04, i-z = -0.11$\pm$0.07, and g-i = 0.60$\pm$0.04, corresponding to a spectral slope of 0.67$\pm$0.40~$\%$/100 nm. The closest match to an asteroid spectral type is with B-type asteroids from the C-complex. We detect variations in the time series photometry of the asteroid with an amplitude of $\sim$0.75, and a double-peaked rotation period of $\sim$1900 s. Assuming a visible albedo of 0.07$\pm$0.03, a density of $\sim$1500 kg/m$^3$, and a calculated absolute magnitude of 30.92$\pm$0.05, we estimate that the asteroid has a diameter of 3.3$\pm$0.7 m and a total mass of $\sim$28,000 kg. Comparing our astrometric orbital solutions to NEOMOD3, the most likely source of \rw is the 3:1 main belt mean motion resonance (77\% probability) followed by the $\nu_6$ resonance (13\% probability), consistent with its organic B-type nature.
\end{abstract}

\section{Introduction}
Imminent impactors provide an opportunity to test the link between meteorites and their parent bodies in the Main Belt (MB) where they likely originated \citep[][]{Granvik2016, Chow2025}. Characterization of imminent impactors using reflected light can be directly compared with fragments that reach the ground as for 2008 TC$_3$ \citep[][]{Jenniskens2009}. The goal of studying imminent impactors is to obtain their spectra while still in space, to collect fragments that reach the ground, and contrast high-resolution lab spectra of the meteorites with the parent body. As of 2025 May, only 11 imminent impactors have been discovered\footnote{\url{https://cneos.jpl.nasa.gov/pi/}, accessed 2025 May 14.} compared to the $\sim$38,000 known near-Earth objects (NEOs) \footnote{\url{https://cneos.jpl.nasa.gov/stats/totals.html}, accessed 2025 May 14.}. There has been recent success with the recovery of fragments from impacting asteroid 2024 BX$_1$ and the study of its meteor trail \citep[][]{Bischoff2024,Spurny2024}, but obtaining the combination of pre-atmospheric entry characterization and subsequent collection and analysis of meteor fragments remains challenging.

The majority of imminent impactors have been found by all-sky asteroid surveys such as the Catalina Sky Survey \citep[CSS,][]{Larson1998} and the \citep[Asteroid Terrestrial-impact Last Alert System, ATLAS,][]{Tonry2018}. While surveys provide information on a imminent impactor's orbit, dedicated follow-up is required to characterize the asteroids before impact. Most of the imminent impactors were discovered  less than 24 h before impact \citep[e.g.,][]{Jenniskens2021,Bischoff2024, Gianotto2025} which provides  a only brief window of time to coordinate multi-wavelength observations of these objects. Rapid-response techniques have been developed to address the characterization of small, close-approaching asteroids, such as the multi-filter streak photometry technique developed by \citep[][]{Bolin2024Streak} to characterize small NEOs 2016 GE$_1$, 2016 CG$_{18}$, and 2016 EV$_{84}$, and which was later used to study Earth impactors, \citep[e.g., 2022 WJ$_1$, 2024 BX$_1$,][]{Devogele2024,Kareta2024,Bolin_3I}and fast-turnaround triggering of spectroscopic observations at 8-m and 10-m class observatories \citep[][]{Bolin2025KY26}. Meteor spectrographs also provide an invaluable tool for the characterization of imminent impactors as they pass through Earth's atmosphere \citep[][]{Spurny2024}.

The imminent impactor \rw was discovered by the CSS on 2024 September 04 \citep[][]{Wierzchos2024RW1}, less than 12 h before its impact near Luzon island in the Philippines on 2024 September 04 16:39 \citep[][]{VeresCBET2025}. This paper describes its observation and characterization using the Astrophysical Research Consortium (ARC) 3.5-m Telescope at Apache Point Observatory (APO). The characterization and analysis of \rw will use the spectrophotometric approach of \citet[][]{Bolin2021LD2,Bolin2022IVO,Bolin2025YR4}. We also report the object's astrometry used to refine its orbit and impact location, and provide its likely source regions within the MB using the NEOMOD3 \citep[][]{Nesvorny2023NEOMOD}.

\section{Observations}
\label{s.obs}
Initially designated as CAQTDL2, \rw was first detected by the CSS 1.5 m telescope  on Mt. Lemmon (Minor Planet Center, MPC, observatory code G96) on 2024 September 04 05:43 UTC. Subsequent observations from both G96 and the Magdalena Ridge Observatory, Socorro 2.4-m telescope \citep[MPC observatory code H01,][]{Pentland2006} refined its orbital solution and increased its impact probability to 1, setting off alerts from the Center for Near Earth Object Studies (CNEOS) Scout system and the European Space Agency's Meerkat system \citep[][]{VeresCBET2025}. They calculated that \rw would impact near the northern Philippines at 18.083657$^{\circ}$N,  123.002342$^{\circ}$ E on 2024 September 16:39 UTC. 

Following the alerts from NEO Scout, we triggered our observations using the ARC 3.5-m telescope's Astrophysical Research Consortium Telescope Imaging Camera (ARCTIC) instrument under observing program JH04 (PI: Ingebretsen) at 2024 September 4 09:58 UTC and ended our observations at 2024 September 04 10:57 UTC. ARCTIC has a 7.85 arcmin x 7.85 arcmin field of view and a 0.228 arcsec/pixel scale when binned in 2 x 2  mode \citep[][]{Huehnerhoff2016}. We conducted photometric time series observations of \rw using the ARCTIC imaging camera with the Sloan Digital Sky Survey (SDSS) g, r, i, and z filters with effective wavelengths 477 nm, 623 nm, 763 nm and 913 nm respectively \citep[][]{Fukugita1996}. At the beginning of the observations \rw was $\approx 500,000 km$ from Earth (Fig.~1), 1.01 au from the Sun, and had a phase angle of $\sim$11.6$^{\circ}$. The seeing in the r-band images ranged from $\sim$1.4-1.8 arcsec and the airmass of the observations increased from 2.6 to 5.3 through the ARC 3.5m observing run.

\begin{figure}
\centering
\includegraphics[width=1.1\linewidth]{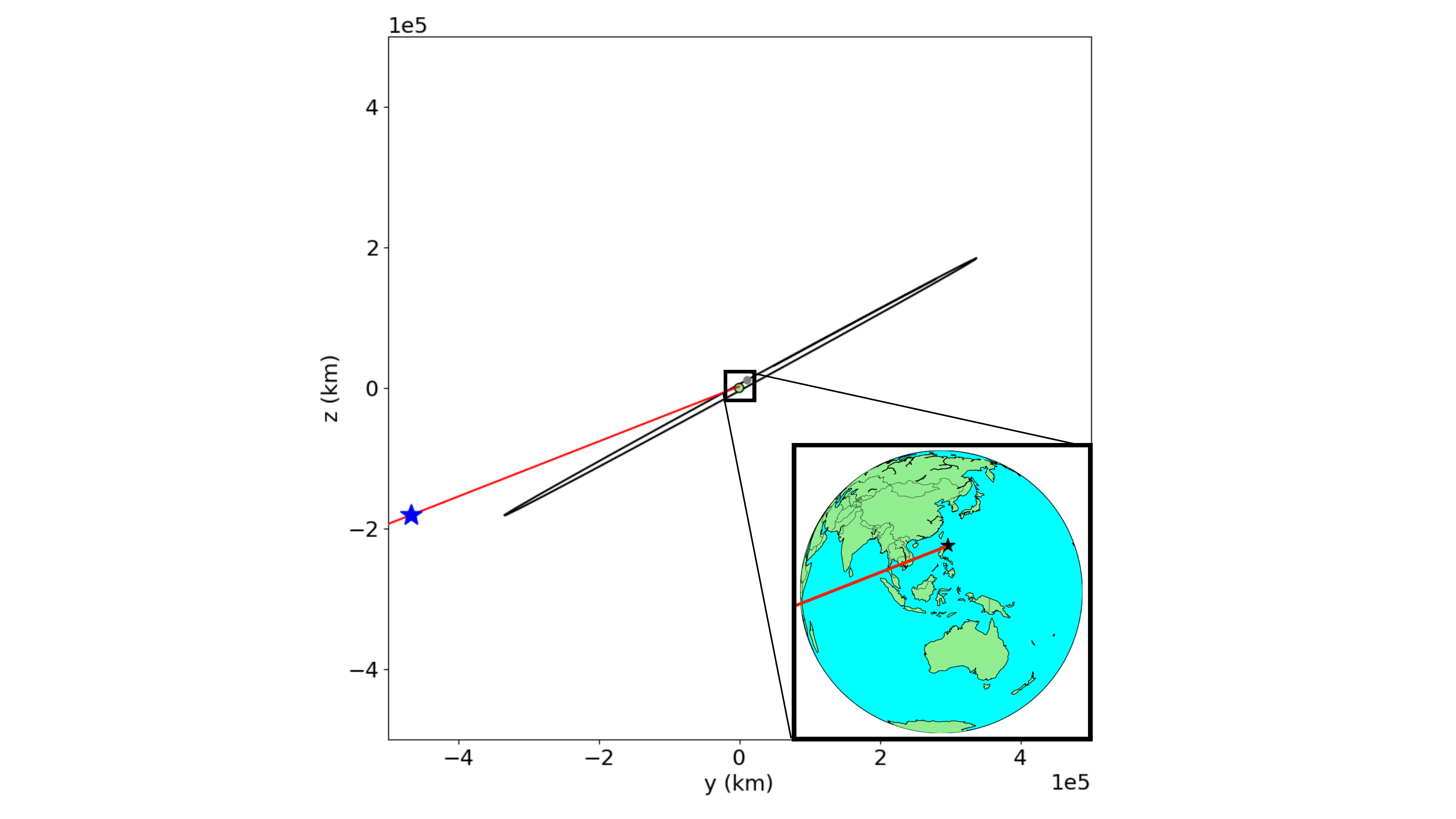}
\caption{The path of \rw is indicated by the red line while the Earth is represented by a green circle drawn to scale. The black curve is the orbit of the Moon with the location of the Moon at the time of observation indicated by a grey circle (not to scale). The insert shows a zoomed-in view of the path over Earth. The final point of impact at 18.083657$^{\circ}$N,  123.002342$^{\circ}$ E is marked by a black star. The blue star indicates the location where we observed \rw before its impact. Both the main plot and insert are in cartesian equatorial coordinates plotted in the YZ plane. The YZ plane has been used in this plot to show the impact location of \rw on the Earth's surface from the reader's point of view.}
    \label{fig:enter-label}
\end{figure}

We initially acquired \rw using 45 second exposures in the SDSS r-band filter tracking sidereally. Once \rw was identified we adjusted the tracking rate to match its rate of motion and began imaging it sequentially in four SDSS filters in a (rgizig) pattern to avoid biasing the time series photometry and improve the measurement of the asteroid's period. All the exposures in all four filters had the same exposure time of 45 seconds. In total we acquired 7 x g-band images, 10 x  r-band images, 6 x i-band images and 6 x z-band images with our last image at 2024 September 04 10:57 UTC. At the end of the night, flat-field and bias frames were taken to calibrate the imaging.

Each image was bias subtracted with a median-combined master bias and then flat-fielded with a median-combined flat for the specific filter. The i and z-band images suffered from significant fringing. To correct for this effect we constructed `fringe-flats' by median combining 45 second exposures of sky with relatively sparse background stars by 2 arcminutes each. Median stacks of the detrended, non-sidereally tracked, g, r, i, and z images are shown in Fig.~2.  The photometry of \rw was measured using an aperture of 4.56 arcsec, and a sky subtraction annulus with an inner radius of 5.70 arcsec, and an outer radius of 6.84 arcsec. The photometry was calibrated using nearby stellar sources from the Pan-STARRS catalogue \citep[][]{Tonry2012,Chambers2016}.

\begin{figure}
\centering
\includegraphics[width=0.6\linewidth]{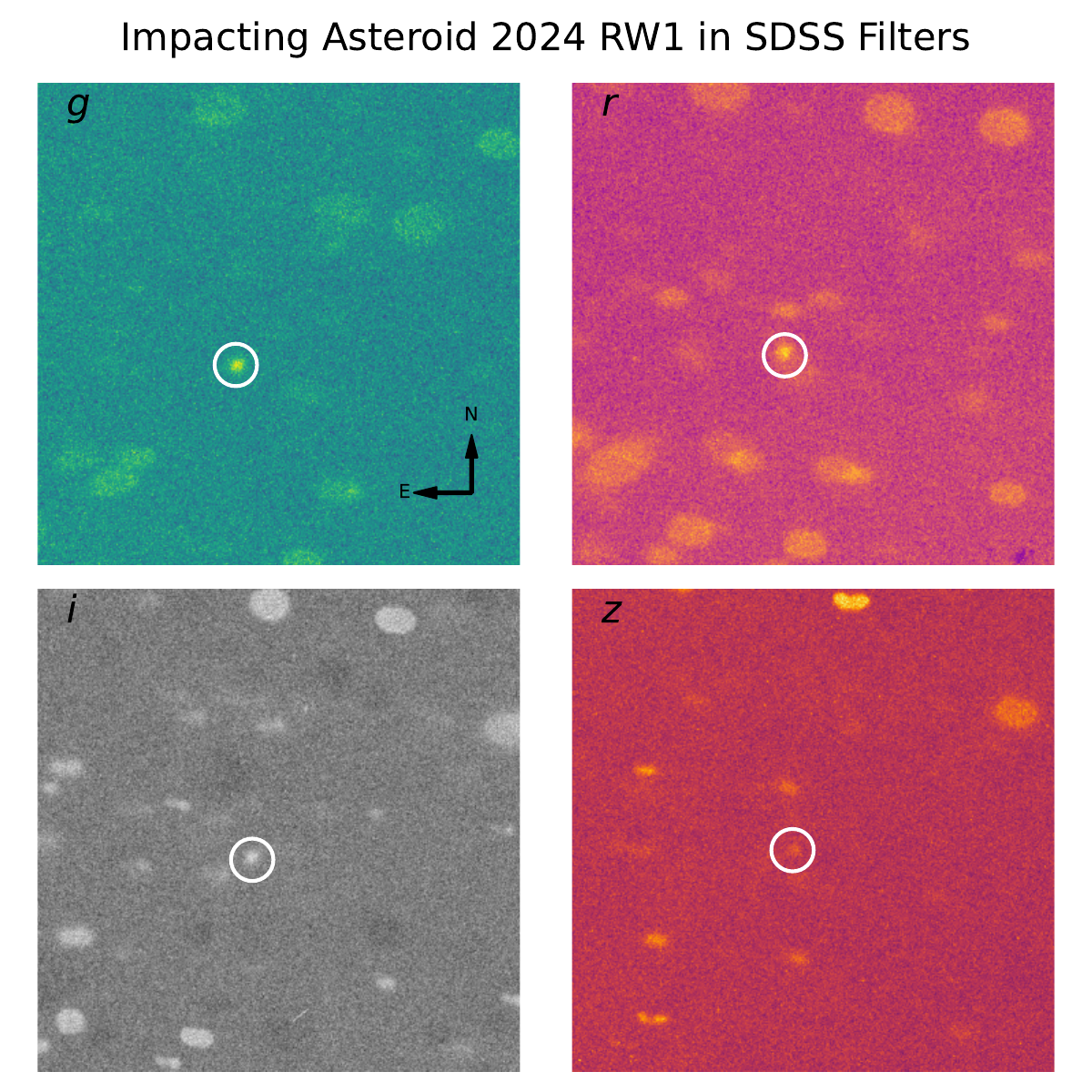}
\caption{\textbf{Top left panel:} a median-combined stack of 7 x 45 s g filter images of \rw encircled in white. \textbf{Top right panel:} a median-combined stack of 10 x 45 s r filter images of \rwns. \textbf{Bottom left panel:} a median-combined stack of 6 x 45 s i filter images of \rwns. \textbf{Bottom right panel:} a median-combined stack of 5 x 45 s \textbf{z} filter images of \rwns. The images were taken with the telescope tracked at the asteroid's apparent rate of motion. The orientation of each image is indicated by the compass rose. Each image stack is 114 arcsec by 114 arcsec.}
\end{figure}

\section{Results}
\subsection{Astrometry}
We measured the astrometry using the Astrometrica software  \citep[][]{Raab2012a} and \textit{Gaia} data release 2 reference stars \citep[][]{Gaia2016,Gaia2018}. The astrometry was submitted to the MPC and appeared in Daily Orbit Update R79 \citep[][]{Williams2024RW1}. In total, 76 observations of \rw were submitted between 2024 September 4 5:43 UTC and 2024 September 4 16:02 UTC which were used to fit a final orbit solution by the Jet Propulsion Laboratory (JPL) Horizons system. The solution from the date 2024 September 09 09:22\footnote{\url{https://ssd.jpl.nasa.gov/tools/sbdb_lookup.html\#/?sstr=2024\%20RW1\&view=OPC}}) is shown in Table~1. The trajectory and final impact location are shown in Fig.~1.


\begin{table}
\centering
\caption{Heliocentric orbital elements of \rw  based on observations collected between 2024 September 4 05:43:41 UTC and 2024 September 4 16:02:23 UTC. The orbital elements are shown for the Julian date (JD) are from JPL Horizons. The solution date is 2024 September 09 09:22. The 1~$\sigma$ uncertainties are given in parentheses.}
\label{t:orbit}
\begin{tabular}{lll}
\hline
Heliocentric Elements&
\\ \hline
Epoch (JD) & 2,460,557.5\\
\hline
Time of perihelion, $T_p$ (JD) & 2,460,601.046693 & $\pm$(5.9304x10$^{-4}$)\\
Semi-major axis, $a$ (au) & 2.507108 & $\pm$(2.039x10$^{-4}$)\\
Eccentricity, $e$ &0.7067746 & $\pm$(2.627x10$^{-5}$)\\
Perihelion, $q$ (au) & 0.73514815 & $\pm$(6.0749x10$^{-6}$)\\
Aphelion, $Q$ (au) & 4.279068 & $\pm$(3.480x10$^{-4}$)\\
Inclination, $i$ ($^{\circ}$) & 0.5280530 & $\pm$(1.9849x10$^{-5}$)\\
Ascending node, $\Omega$ ($^{\circ}$) & 162.4574635 & $\pm$(1.1943x10$^{-5}$)\\
Argument of perihelion, $\omega$ ($^{\circ}$) & 249.6223582 & $\pm$(4.0094x10$^{-5}$)\\
Mean Anomaly, $M$ ($^{\circ}$) & 349.18816 & $\pm$(1.466x10$^{-3}$)\\
\hline
\end{tabular}
\end{table}

\subsection{Photometry and spectral classification}
\label{sec:photo}
The median brightness of \rw in the four filters was g = 19.41 $\pm$ 0.03, r = 18.94 $\pm$ 0.03, i = 18.15 $\pm$ 0.03, z = 18.92 $\pm$ 0.07 and its color indices are  g-r = 0.47$\pm$0.04, r-i = 0.13$\pm$0.04, and i-z = -0.11$\pm$0.07.  Its colors are comparable to the Sun in g, r, and i bands (g$\mathrm{_{sun}}$-r$\mathrm{_{sun}}$ = 0.44, r$\mathrm{_{sun}}$-i$\mathrm{_{sun}}$ = 0.11, but it is bluer than the Sun in i-z \citep[i$\mathrm{_{sun}}$-z$\mathrm{_{sun}}$  = 0.3, ][]{Willmer2018}. The g-i color and spectra slope are 0.60$\pm$0.04 and 0.67$\pm0.40~\%$/100 nm. All error estimates are 1-sigma.

Defining a$^{*}$ as a$^{*}$ = (0.89 (g-r)) + (0.45 (r-i)) - 0.57  \citep[][]{Ivezic2001},  we find a$^{*}$ = -0.09$\pm$0.06 (Fig.~3). \rw has a$^{*}$ and i-z that are similar to C-types which have a$^{*}$ $\simeq$ -0.1 and i-z $\simeq$ 0.02 but it is significantly bluer than S-types and V-types which have an a$^{*}$ $\simeq$ 0.13-0.15 (Fig.~4). \rw has a relatively small (in magnitude) i-z = -0.11$\pm$0.07 compared to V-types, which have a large (in magnitude) i-z $\simeq$ -0.5 due to the presence of strong silicate absorption features at $\sim$1 $\mu$m \citep[][]{DeMeo2013aa}.

\begin{figure}
\centering
\includegraphics[width=0.7\linewidth]{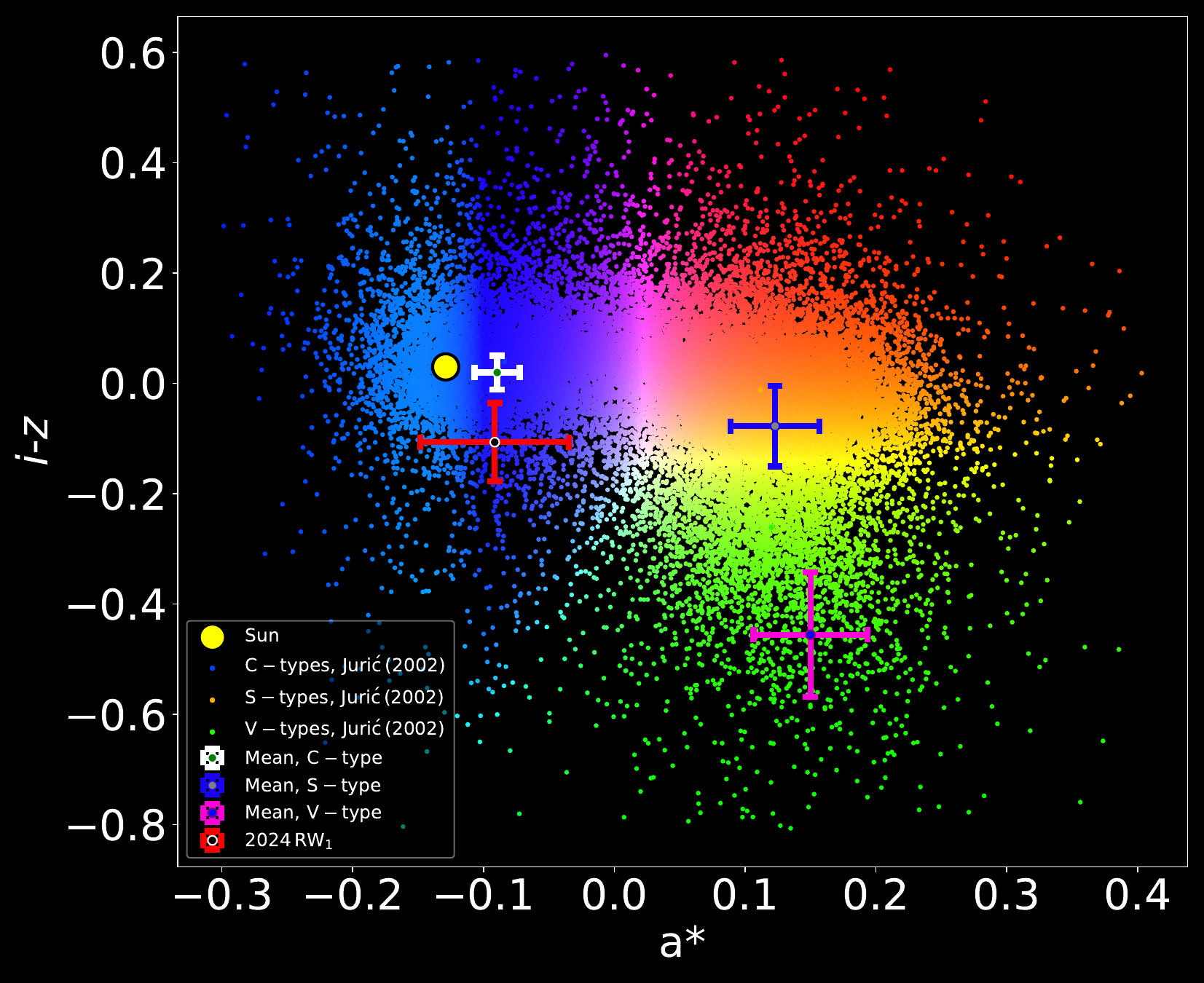}
\caption{SDSS a$^{*}$ vs. i-z color indices of \rw plotted with a$^{*}$ vs. i-z colors of C (blue), S (red) and V (green) type asteroids  \citep{Ivezic2001,Juric2002}. The colorization scheme of data points as a function of a$^{*}$ and i-z is adapted from \citet[][]{Ivezic2002}. The a$^{*}$ and $i$-$z$ range of average S, V, and C-type asteroids are shown computed from the average spectra from \citet[][]{DeMeo2009}.}
\end{figure}

We computed \rw's spectral reflectance in each filter by dividing its per-filter flux by the flux of a solar-type star in the corresponding filter and normalized to the spectral reflectance at 550 nm (Fig.~4). The spectrum of \rw is similar to the spectra of B- and C-type asteroids \citep[][]{DeMeo2009} with a relatively flat slope between 477 nm and 763 nm and a slight decrease in reflectance near 913 nm. We then computed the $\chi^2$ statistic between the \rw's spectrum and the asteroid taxonomic catalogue spectra of \citet[][]{DeMeo2009}. We used 22 average catalogue spectra including S-complex asteroid types (S, Sa, Sq, Sr, Sv, Q),  C-complex types (B, C, Cg, Cgh), X-complex types (X, Xc, Xe, Xk, Xn), and other assorted types (A, D, K, L, O, R, V). The closest match is to B-type asteroids with a reduced $\chi^2$ of $\sim$1 followed by C-type asteroids with a reduced $\chi^2$ of $\sim$2.6. By comparison, the $\chi^2$ with the S-types is $\sim$31.3 and V-types is $\sim$29.6.

\begin{figure}
\centering
\includegraphics[width=0.82\linewidth]{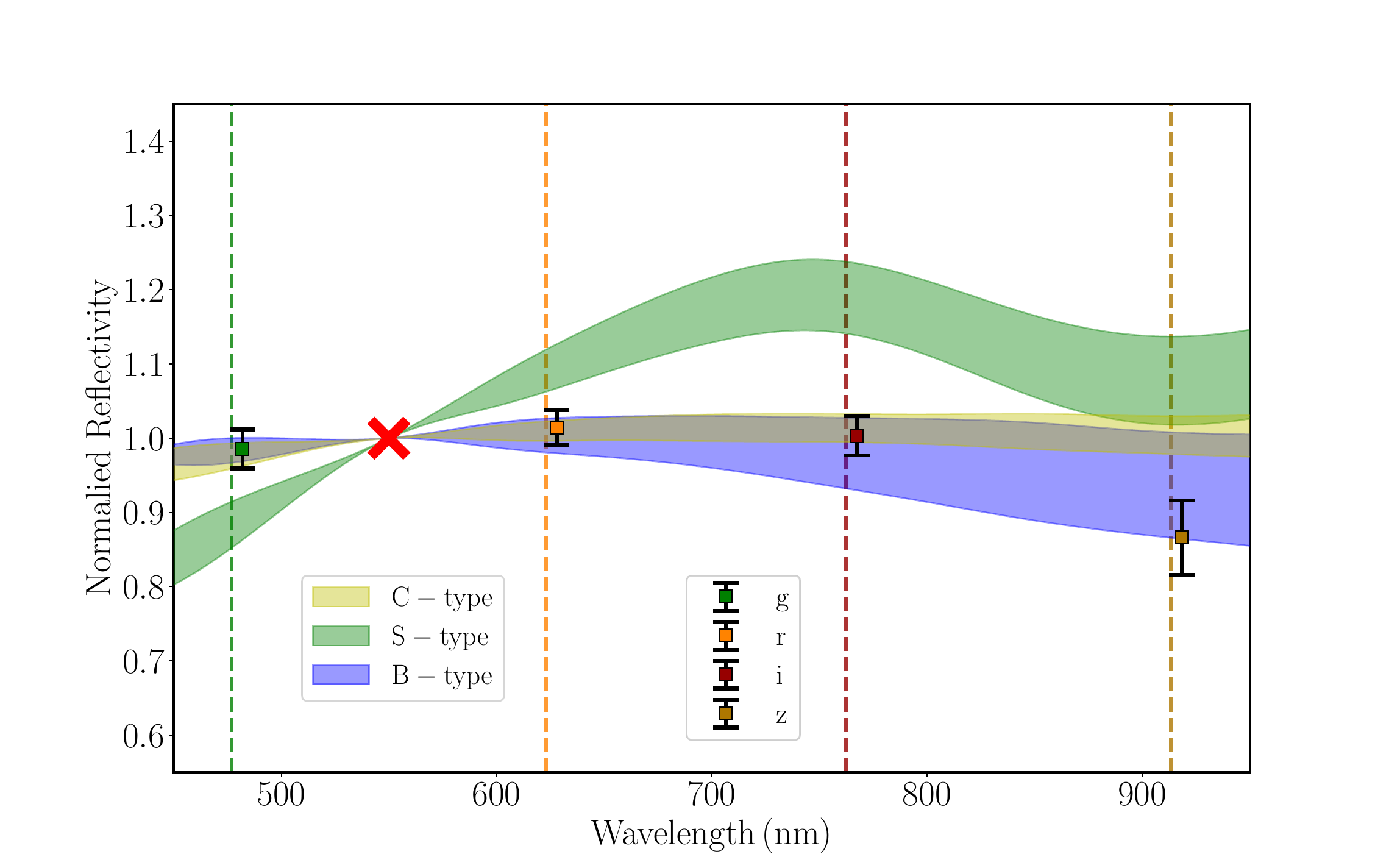}
\caption{Reflectance spectrophotometry of \rw consisting of g, r, i, and z observations of \rw on 2024 September 4 UTC. The $\lambda_{\mathrm{eff}}$ locations of the g, r, i, and z filters have been plotted as vertical dashed lines. The data points for the normalized reflectivity of \rw have been offset slightly from their location in the wavelength direction. The error bars on the data points correspond to 1$\sigma$ uncertainty. The spectrum has been normalized to unity at 550 nm, as indicated by the red X. The spectral range of C, S and B-type asteroids from the Bus-DeMeo asteroid taxonomic catalog \citep[][]{DeMeo2009} are over-plotted with the B-type spectrum most closely resembling the spectra of \rw.}
\end{figure}
We estimate the absolute magnitude, $H$, using our g and r observations of \rwns. First we convert the our g and r magnitudes to a V magnitude using a transformation equation from \citet[][]{Jester_2005}, 
\begin{equation}
    \label{V_transform}
    V = g - 0.59(g - r) - 0.01.
\end{equation}
We calculate a V magnitude of $19.12 \pm 0.05$. Then we can calculate the absolute magnitude with the following equation
\begin{equation}
\label{eqn.brightness}
H = V - 5\, \mathrm{log_{10}}(r_h \Delta) +2.5\,\mathrm{log_{10}}\left[ (1 - G)\,\Phi_1(\alpha) + G\,\Phi_2(\alpha) \right ]
\end{equation}
taken from \citet[][]{Bowell1988}. Here, $r_h$ is the heliocentric distance of 1.011 au, $\Delta$ is the geocentric distance of 0.003 au, $\alpha$ is the phase angle of 10.872$^{\circ}$ that \rw had during our observations, $G$ is the phase coefficient which we adopt the value of 0.03, the average value for C-type asteroids \citep[][]{Veres2015}, and $\Phi_1(\alpha)$ and $\Phi_2(\alpha)$ are the basis functions normalized at $\alpha$ = 0$^\circ$ described in \citep[][]{Bowell1988}. Using the above equation, we estimate that $H$ = 30.92$\pm$0.05, though our uncertainty on $H$ is underestimated due to the unknown phase function of \rwns.

\subsection{Rotational lightcurve}
\label{s.lightcurve}

We searched for periodic variations of the r-band lightcurve of \rw presented in Table~2. The amplitude of the variations are $\sim$0.75, significantly larger than the $\sim$0.5-0.1 uncertainties of the individual data points (Fig.~5). We used a similar approach as \citet[][]{Bolin2025PT5} in testing the periodicity of the lightcurve by using the Lomb-Scargle (L-S) periodogram \citep[][]{Lomb1976} (middle panel of Fig.~5) and obtained a single-peaked lightcurve of $\sim$976 s. We take a boot strap approach to estimate the lightcurve period and its uncertainty for the r-band data, similar to \citet[][]{Bolin2020CD3} and \citet[][]{Purdum2021}, by removing $\sqrt{N}$ data points from the time series lightcurve and repeating our L-S periodogram estimation of the lightcurve period 10,000 times, resulting in a central value and 1-sigma uncertainty estimate of $\sim$979$\pm$ $\sim$59 s. We adopted the approach of \citet[][]{Bolin2020HST} by using phase dispersion minimization (PDM) analysis \citep[][]{Stellingwerf1978} to check the L-S results with an independent method, resulting in a local minima of $\sim$950 s, comparable with the Lomb-Scargle period estimate. Taken together, the L-S and PDM methods imply that \rw has a double-peaked rotation period on the order of $\sim$1900 s.

\begin{figure}
\centering
\includegraphics[scale=0.35]{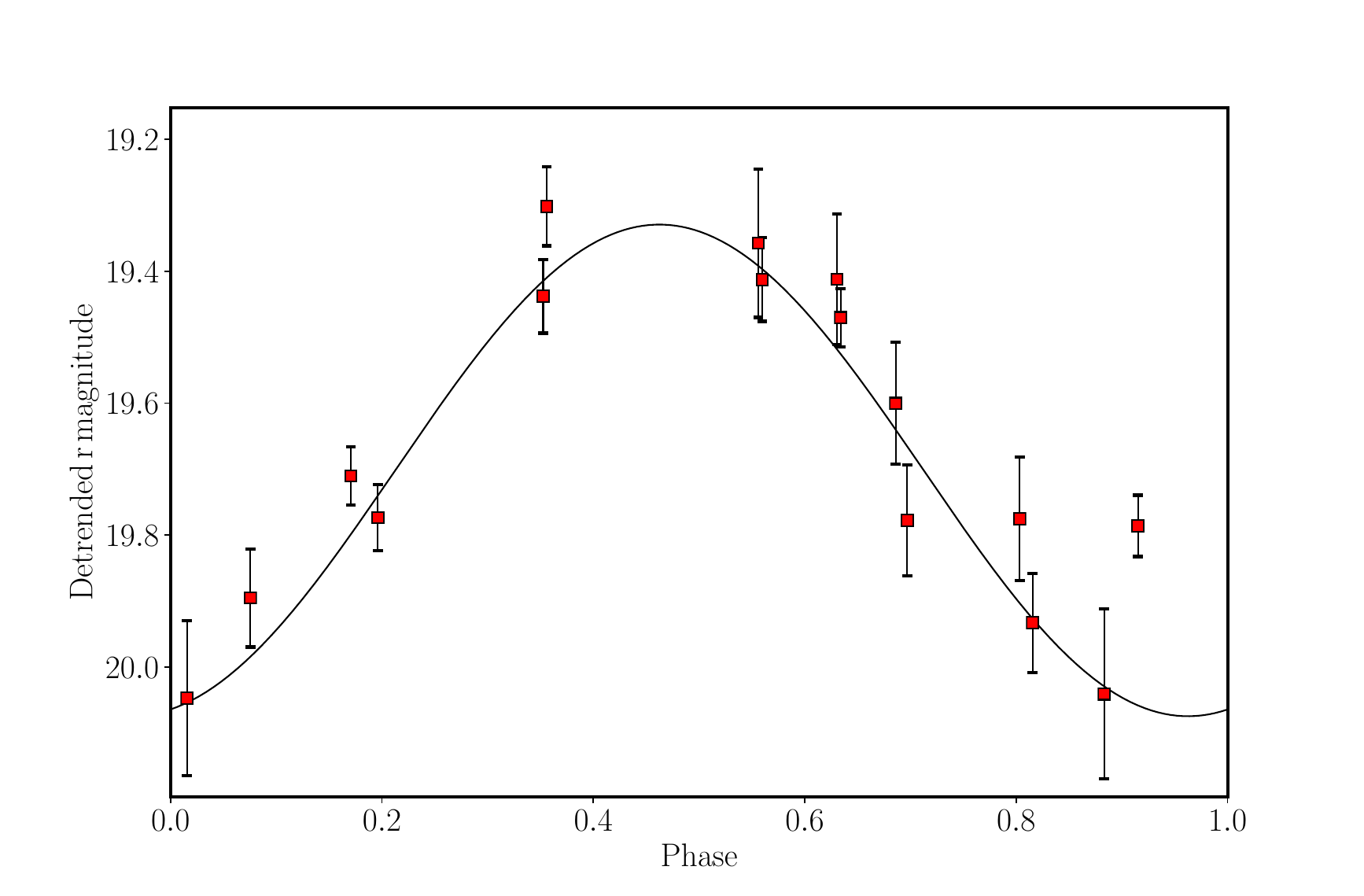}
\includegraphics[scale=0.35]{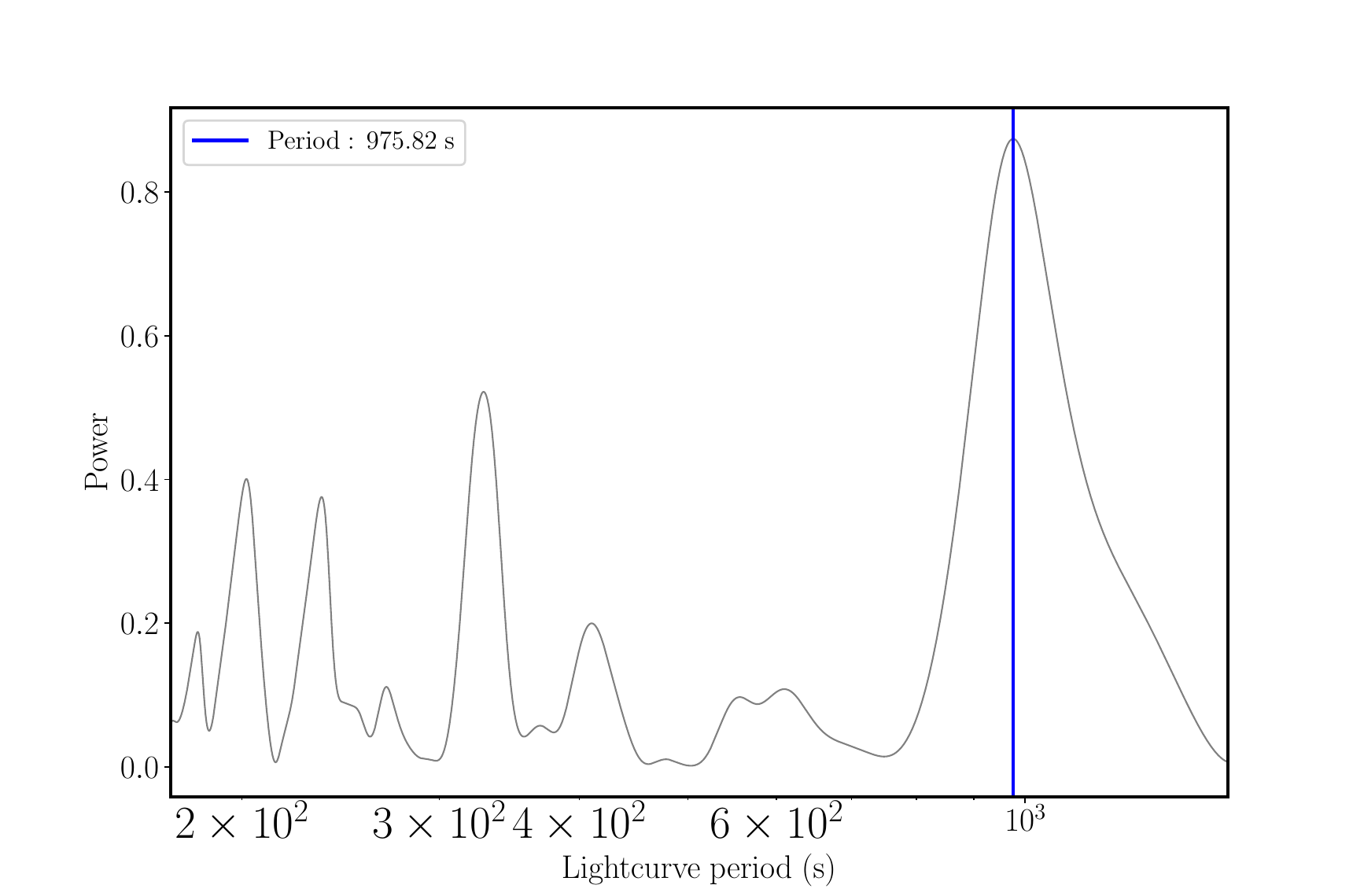}
\includegraphics[scale=0.35]{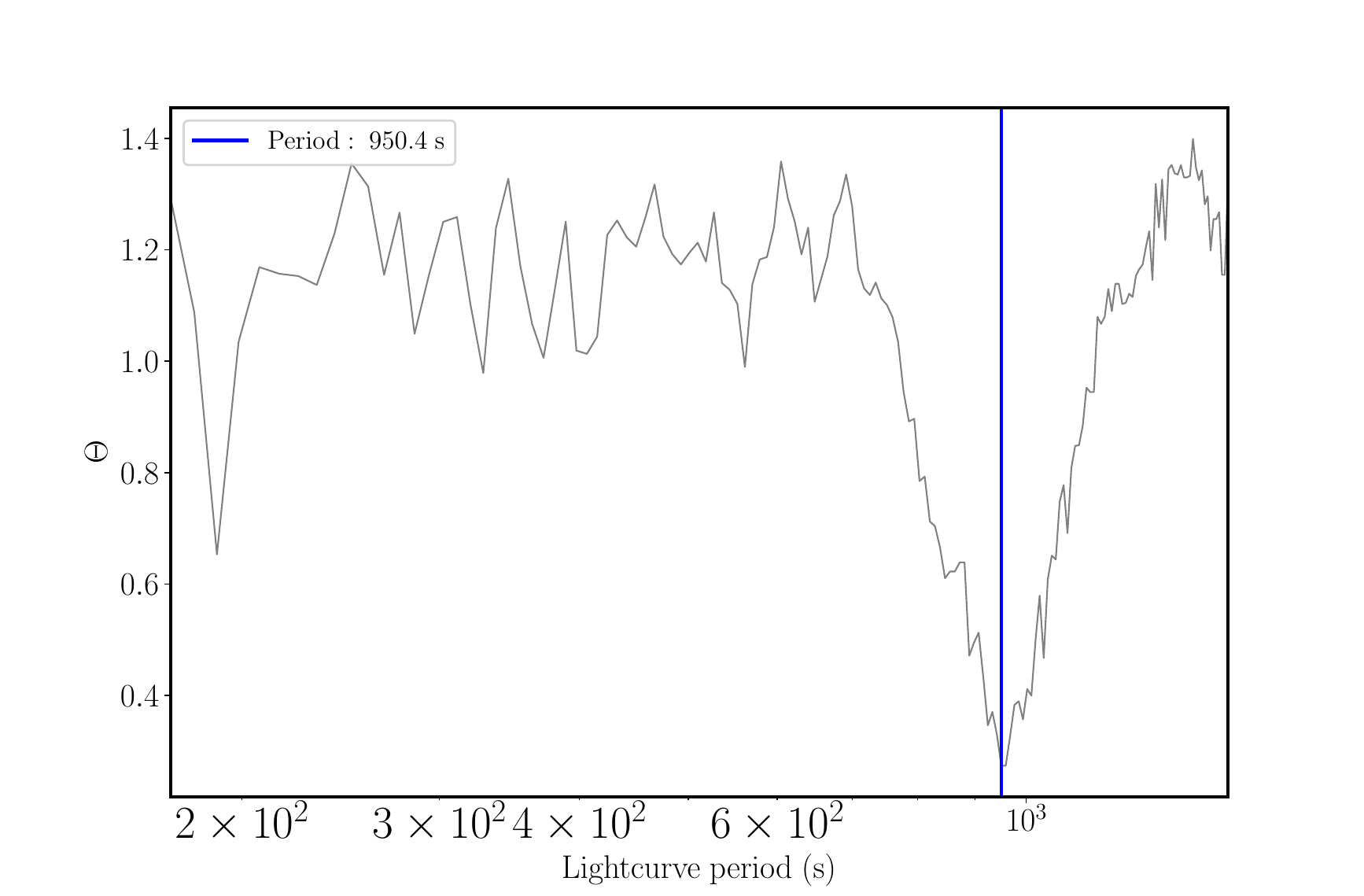}
\caption{
 \textbf{Top panel:} phased r filter lightcurve data of \rw using a single-peaked lightcurve period of $\sim$950 s.
\textbf{Middle panel:} Lomb-Scargle periodogram of lightcurve period vs. spectral power \citep[][]{Lomb1976} for the r-band images. A peak in the power is located at a single-peaked lightcurve period of 975.9, implying a double-peaked lightcurve of 1951.8 s}.
\textbf{Bottom panel:} phase dispersion minimization analysis of lightcurve rotation period vs. $\Theta$ metric \citep[][]{Stellingwerf1978}. The $\Theta$ metric is minimized at a single-peaked rotation period of 950.4 s implying a double-peaked rotation period of 1900.8 s.
\end{figure}

\section{Discussion and Conclusions}

We can use the spectral and orbital information of \rw to estimate its visible albedo, $p_v$, and its size and MB origin. The spectrum of \rw most closely matched the spectra of B-type asteroids, which are carbonaceous and have an average visible albedo of 0.07$\pm$0.03 \citep[][]{AliLagoa2013}. We can then calculate the object's diameter, $\mathrm{D = \frac{1329}{\sqrt{p_v}}10^{-\frac{H}{5}}}$, \citep[][]{Harris2002} using the absolute magnitude, $H$, of 30.92$\pm$0.05 from Section~3.2, finding that D = 3.3$\pm$0.7 m. Assuming a density of $\sim$1500 kg/m$^3$ typical of B-type asteroids \citep[][]{Hanus2017density}, \rw's mass was $\sim$28,000 kg. Lastly, we can estimate a rough shape for \rw using its observed 0.75 magnitude light curve amplitude. Assuming a triaxial shape with a:b:c and a$\simeq$c, then $b/a \; = \; 10^{0.4 A}$ \citep[][]{Binzel1989}. Using our observed amplitude of $A$ = 0.75, we find a $b/a\sim$2, implying that \rw is elongated similar to other minor bodies \citep[][]{Hudson1999,Mann2007,Bolin2018}.

We determined the likely MB source of \rw from its $a$, $e$, $i$, and H, and the NEOMOD3 NEO population model \citep[][]{Nesvorny2023NEOMOD, Nesvorny2024NEOMOD3} and found that it is most likely from the 3:1 mean motion resonance with Jupiter, at a $\sim$77$\%$ probability, followed by the $\nu_6$ resonance, at a $\sim$13$\%$ probability, and the 7:3 mean motion resonance, at a $\sim$6$\%$ probability. Therefore, it seems that \rw is from the border between the inner MB and outer MB where several primitive asteroid families are located \citep[][]{Walsh2013, Delbo2017}. Current and future asteroid \citep[e.g.,][]{Larson1998,Denneau2013,Tonry2018,Masiero2024} and general-purpose all-sky surveys \citep[e.g., ][]{Whidden2019,Vera2020} may result in the detection of additional imminent impactors providing opportunities for the characterization of the Earth-impactor population.

\section*{acknowledgments}

Based on observations obtained with the Apache Point Observatory 3.5-meter telescope, which is owned and operated by the Astrophysical Research Consortium.

\facility{ARC} 

\bibliographystyle{aasjournal}
\bibliography{neobib_fixed}

\begin{thebibliography}{}
\expandafter\ifx\csname natexlab\endcsname\relax\def\natexlab#1{#1}\fi
\providecommand{\url}[1]{\href{#1}{#1}}
\providecommand{\dodoi}[1]{doi:~\href{http://doi.org/#1}{\nolinkurl{#1}}}
\providecommand{\doeprint}[1]{\href{http://ascl.net/#1}{\nolinkurl{http://ascl.net/#1}}}
\providecommand{\doarXiv}[1]{\href{https://arxiv.org/abs/#1}{\nolinkurl{https://arxiv.org/abs/#1}}}

\bibitem[{{Al{\'\i}-Lagoa} {et~al.}(2013){Al{\'\i}-Lagoa}, {de Le{\'o}n},
  {Licandro}, {Delb{\'o}}, {Campins}, {Pinilla-Alonso}, \&
  {Kelley}}]{AliLagoa2013}
{Al{\'\i}-Lagoa}, V., {de Le{\'o}n}, J., {Licandro}, J., {et~al.} 2013, \aap,
  554, A71, \dodoi{10.1051/0004-6361/201220680}

\bibitem[{{Binzel} {et~al.}(1989){Binzel}, {Farinella}, {Zappal\`a}, \&
  {Cellino}}]{Binzel1989}
{Binzel}, R.~P., {Farinella}, P., {Zappal\`a}, V., \& {Cellino}, A. 1989, in
  Asteroids II, ed. R.~P. {Binzel}, T.~{Gehrels}, \& M.~S. {Matthews}, 416--441

\bibitem[{{Bischoff} {et~al.}(2024){Bischoff}, {Patzek}, {Barrat}, {Berndt},
  {Busemann}, {Degering}, {Di Rocco}, {Ek}, {Harries}, {Godinho}, {Heinlein},
  {Kriele}, {Krietsch}, {Maden}, {Marchhart}, {Marshal}, {Martschini},
  {Merchel}, {M{\"o}ller}, {Pack}, {Raab}, {Reitze}, {Rendtel},
  {R{\"u}fenacht}, {Sachs}, {Sch{\"o}nb{\"a}chler}, {Schuppisser}, {Weber},
  {Wieser}, \& {Wimmer}}]{Bischoff2024}
{Bischoff}, A., {Patzek}, M., {Barrat}, J.-A., {et~al.} 2024, \maps, 59, 2660,
  \dodoi{10.1111/maps.14245}

\bibitem[{{Bolin} {et~al.}(2025){Bolin}, {Denneau}, {Abron}, {Jedicke},
  {Chiboucas}, {Ingebretsen}, \& {Lemaux}}]{Bolin2025PT5}
{Bolin}, B.~T., {Denneau}, L., {Abron}, L.-M., {et~al.} 2025, \apjl, 978, L37,
  \dodoi{10.3847/2041-8213/ada1d0}

\bibitem[{{Bolin} {et~al.}(2024){Bolin}, {Ghosal}, \&
  {Jedicke}}]{Bolin2024Streak}
{Bolin}, B.~T., {Ghosal}, M., \& {Jedicke}, R. 2024, \mnras, 527, 1633,
  \dodoi{10.1093/mnras/stad3227}

\bibitem[{{Bolin} \& {Lisse}(2020)}]{Bolin2020HST}
{Bolin}, B.~T., \& {Lisse}, C.~M. 2020, \mnras, 497, 4031,
  \dodoi{10.1093/mnras/staa2192}

\bibitem[{{Bolin} {et~al.}(2018){Bolin}, {Weaver}, {Fernandez}, {Lisse},
  {Huppenkothen}, {Jones}, {Juri{\'c}}, {Moeyens}, {Schambeau}, {Slater},
  {Ivezi{\'c}}, \& {Connolly}}]{Bolin2018}
{Bolin}, B.~T., {Weaver}, H.~A., {Fernandez}, Y.~R., {et~al.} 2018, \apjl, 852,
  L2, \dodoi{10.3847/2041-8213/aaa0c9}

\bibitem[{{Bolin} {et~al.}(2020){Bolin}, {Fremling}, {Holt}, {Hankins},
  {Ahumada}, {Anand}, {Bhalerao}, {Burdge}, {Copperwheat}, {Coughlin},
  {Deshmukh}, {De}, {Kasliwal}, {Morbidelli}, {Purdum}, {Quimby}, {Bodewits},
  {Chang}, {Ip}, {Hsu}, {Laher}, {Lin}, {Lisse}, {Masci}, {Ngeow}, {Tan},
  {Zhai}, {Burruss}, {Dekany}, {Delacroix}, {Duev}, {Graham}, {Hale},
  {Kulkarni}, {Kupfer}, {Mahabal}, {Mr{\'o}z}, {Neill}, {Riddle}, {Rodriguez},
  {Smith}, {Soumagnac}, {Walters}, {Yan}, \& {Zolkower}}]{Bolin2020CD3}
{Bolin}, B.~T., {Fremling}, C., {Holt}, T.~R., {et~al.} 2020, \apjl, 900, L45,
  \dodoi{10.3847/2041-8213/abae69}

\bibitem[{{Bolin} {et~al.}(2021){Bolin}, {Fernandez}, {Lisse}, {Holt}, {Lin},
  {Purdum}, {Deshmukh}, {Bauer}, {Bellm}, {Bodewits}, {Burdge}, {Carey},
  {Copperwheat}, {Helou}, {Ho}, {Horner}, {van Roestel}, {Bhalerao}, {Chang},
  {Chen}, {Hsu}, {Ip}, {Kasliwal}, {Masci}, {Ngeow}, {Quimby}, {Burruss},
  {Coughlin}, {Dekany}, {Delacroix}, {Drake}, {Duev}, {Graham}, {Hale},
  {Kupfer}, {Laher}, {Mahabal}, {Mr{\'o}z}, {Neill}, {Riddle}, {Rodriguez},
  {Smith}, {Soumagnac}, {Walters}, {Yan}, \& {Zolkower}}]{Bolin2021LD2}
{Bolin}, B.~T., {Fernandez}, Y.~R., {Lisse}, C.~M., {et~al.} 2021, \aj, 161,
  116, \dodoi{10.3847/1538-3881/abd94b}

\bibitem[{{Bolin} {et~al.}(2022){Bolin}, {Ahumada}, {van Dokkum}, {Fremling},
  {Granvik}, {Hardegree-Ullman}, {Harikane}, {Purdum}, {Serabyn}, {Southworth},
  \& {Zhai}}]{Bolin2022IVO}
{Bolin}, B.~T., {Ahumada}, T., {van Dokkum}, P., {et~al.} 2022, \mnras, 517,
  L49, \dodoi{10.1093/mnrasl/slac089}

\bibitem[{Bolin {et~al.}(2025)Bolin, Belyakov, Fremling, Graham, Abdelaziz,
  Elhosseiny, Gray, Ingebretsen, Jewett, Lisse, Karpov, Kilic, Mašek, Molham,
  Roderick, Takey, Abron, Coughlin, Hsieh, Noll, \& Wong}]{Bolin_3I}
Bolin, B.~T., Belyakov, M., Fremling, C., {et~al.} 2025, Monthly Notices of the
  Royal Astronomical Society: Letters, 542, L139,
  \dodoi{10.1093/mnrasl/slaf078}

\bibitem[{{Bolin} {et~al.}(2025{\natexlab{a}}){Bolin}, {Fremling}, {Belyakov},
  {Beniyama}, {Delbo}, {Jedicke}, {Wong}, {Abron}, {Noll}, \&
  Stephens}]{Bolin2025KY26}
{Bolin}, B.~T., {Fremling}, C., {Belyakov}, M., {et~al.} 2025{\natexlab{a}},
  \aj, 169, 7pp, \dodoi{10.3847/1538-3881/adccbe}

\bibitem[{{Bolin} {et~al.}(2025{\natexlab{b}}){Bolin}, {Hanu{\v{s}}},
  {Denneau}, {Bonamico}, {Abron}, {Delbo}, {{\v{D}}urech}, {Jedicke}, {Alcorn},
  {Cikota}, {Panda}, \& {Reggiani}}]{Bolin2025YR4}
{Bolin}, B.~T., {Hanu{\v{s}}}, J., {Denneau}, L., {et~al.} 2025{\natexlab{b}},
  \apjl, 984, L25, \dodoi{10.3847/2041-8213/adc910}

\bibitem[{{Bowell} {et~al.}(1988){Bowell}, {Hapke}, {Domingue}, {Lumme},
  {Peltoniemi}, \& {Harris}}]{Bowell1988}
{Bowell}, E., {Hapke}, B., {Domingue}, D., {et~al.} 1988, Asteroids II, 399

\bibitem[{{Chambers} {et~al.}(2016){Chambers}, {Magnier}, {Metcalfe},
  {Flewelling}, {Huber}, {Waters}, {Denneau}, {Draper}, {Farrow}, {Finkbeiner},
  {Holmberg}, {Koppenhoefer}, {Price}, {Saglia}, {Schlafly}, {Smartt},
  {Sweeney}, {Wainscoat}, {Burgett}, {Grav}, {Heasley}, {Hodapp}, {Jedicke},
  {Kaiser}, {Kudritzki}, {Luppino}, {Lupton}, {Monet}, {Morgan}, {Onaka},
  {Stubbs}, {Tonry}, {Banados}, {Bell}, {Bender}, {Bernard}, {Botticella},
  {Casertano}, {Chastel}, {Chen}, {Chen}, {Cole}, {Deacon}, {Frenk},
  {Fitzsimmons}, {Gezari}, {Goessl}, {Goggia}, {Goldman}, {Grebel}, {Hambly},
  {Hasinger}, {Heavens}, {Heckman}, {Henderson}, {Henning}, {Holman}, {Hopp},
  {Ip}, {Isani}, {Keyes}, {Koekemoer}, {Kotak}, {Long}, {Lucey}, {Liu},
  {Martin}, {McLean}, {Morganson}, {Murphy}, {Nieto-Santisteban}, {Norberg},
  {Peacock}, {Pier}, {Postman}, {Primak}, {Rae}, {Rest}, {Riess}, {Riffeser},
  {Rix}, {Roser}, {Schilbach}, {Schultz}, {Scolnic}, {Szalay}, {Seitz},
  {Shiao}, {Small}, {Smith}, {Soderblom}, {Taylor}, {Thakar}, {Thiel},
  {Thilker}, {Urata}, {Valenti}, {Walter}, {Watters}, {Werner}, {White},
  {Wood-Vasey}, \& {Wyse}}]{Chambers2016}
{Chambers}, K.~C., {Magnier}, E.~A., {Metcalfe}, N., {et~al.} 2016, ArXiv
  e-prints.
\newblock \doarXiv{1612.05560}

\bibitem[{{Chow} \& {Brown}(2025)}]{Chow2025}
{Chow}, I., \& {Brown}, P.~G. 2025, \icarus, 429, 116444,
  \dodoi{10.1016/j.icarus.2024.116444}

\bibitem[{{Delbo} {et~al.}(2017){Delbo}, {Walsh}, {Bolin}, {Avdellidou}, \&
  {Morbidelli}}]{Delbo2017}
{Delbo}, M., {Walsh}, K., {Bolin}, B., {Avdellidou}, C., \& {Morbidelli}, A.
  2017, Science, 357, 1026, \dodoi{10.1126/science.aam6036}

\bibitem[{{DeMeo} {et~al.}(2009){DeMeo}, {Binzel}, {Slivan}, \&
  {Bus}}]{DeMeo2009}
{DeMeo}, F.~E., {Binzel}, R.~P., {Slivan}, S.~M., \& {Bus}, S.~J. 2009,
  \icarus, 202, 160, \dodoi{10.1016/j.icarus.2009.02.005}

\bibitem[{{DeMeo} \& {Carry}(2013)}]{DeMeo2013aa}
{DeMeo}, F.~E., \& {Carry}, B. 2013, \icarus, 226, 723,
  \dodoi{10.1016/j.icarus.2013.06.027}

\bibitem[{{Denneau} {et~al.}(2013){Denneau}, {Jedicke}, {Grav}, {Granvik},
  {Kubica}, {Milani}, {Vere{\v s}}, {Wainscoat}, {Chang}, {Pierfederici},
  {Kaiser}, {Chambers}, {Heasley}, {Magnier}, {Price}, {Myers}, {Kleyna},
  {Hsieh}, {Farnocchia}, {Waters}, {Sweeney}, {Green}, {Bolin}, {Burgett},
  {Morgan}, {Tonry}, {Hodapp}, {Chastel}, {Chesley}, {Fitzsimmons}, {Holman},
  {Spahr}, {Tholen}, {Williams}, {Abe}, {Armstrong}, {Bressi}, {Holmes},
  {Lister}, {McMillan}, {Micheli}, {Ryan}, {Ryan}, \& {Scotti}}]{Denneau2013}
{Denneau}, L., {Jedicke}, R., {Grav}, T., {et~al.} 2013, \pasp, 125, 357,
  \dodoi{10.1086/670337}

\bibitem[{{Devog{\`e}le} {et~al.}(2024){Devog{\`e}le}, {Buzzi}, {Micheli},
  {Cano}, {Conversi}, {Jehin}, {Ferrais}, {Oca{\~n}a}, {F{\"o}hring}, {Drury},
  {Benkhaldoun}, \& {Jenniskens}}]{Devogele2024}
{Devog{\`e}le}, M., {Buzzi}, L., {Micheli}, M., {et~al.} 2024, \aap, 689, A211,
  \dodoi{10.1051/0004-6361/202450263}

\bibitem[{{Fukugita} {et~al.}(1996){Fukugita}, {Ichikawa}, {Gunn}, {Doi},
  {Shimasaku}, \& {Schneider}}]{Fukugita1996}
{Fukugita}, M., {Ichikawa}, T., {Gunn}, J.~E., {et~al.} 1996, \aj, 111, 1748,
  \dodoi{10.1086/117915}

\bibitem[{{Gaia Collaboration} {et~al.}(2016){Gaia Collaboration}, {Prusti},
  {de Bruijne}, {Brown}, {Vallenari}, {Babusiaux}, {Bailer-Jones}, {Bastian},
  {Biermann}, {Evans}, {Eyer}, {Jansen}, {Jordi}, {Klioner}, {Lammers},
  {Lindegren}, {Luri}, {Mignard}, {Milligan}, {Panem}, {Poinsignon},
  {Pourbaix}, {Randich}, {Sarri}, {Sartoretti}, {Siddiqui}, {Soubiran},
  {Valette}, {van Leeuwen}, {Walton}, {Aerts}, {Arenou}, {Cropper}, {Drimmel},
  {H{\o}g}, {Katz}, {Lattanzi}, {O'Mullane}, {Grebel}, {Holland}, {Huc},
  {Passot}, {Bramante}, {Cacciari}, {Casta{\~n}eda}, {Chaoul}, {Cheek}, {De
  Angeli}, {Fabricius}, {Guerra}, {Hern{\'a}ndez}, {Jean-Antoine-Piccolo},
  {Masana}, {Messineo}, {Mowlavi}, {Nienartowicz}, {Ord{\'o}{\~n}ez-Blanco},
  {Panuzzo}, {Portell}, {Richards}, {Riello}, {Seabroke}, {Tanga},
  {Th{\'e}venin}, {Torra}, {Els}, {Gracia-Abril}, {Comoretto},
  {Garcia-Reinaldos}, {Lock}, {Mercier}, {Altmann}, {Andrae}, {Astraatmadja},
  {Bellas-Velidis}, {Benson}, {Berthier}, {Blomme}, {Busso}, {Carry},
  {Cellino}, {Clementini}, {Cowell}, {Creevey}, {Cuypers}, {Davidson}, {De
  Ridder}, {de Torres}, {Delchambre}, {Dell'Oro}, {Ducourant}, {Fr{\'e}mat},
  {Garc{\'\i}a-Torres}, {Gosset}, {Halbwachs}, {Hambly}, {Harrison}, {Hauser},
  {Hestroffer}, {Hodgkin}, {Huckle}, {Hutton}, {Jasniewicz}, {Jordan},
  {Kontizas}, {Korn}, {Lanzafame}, {Manteiga}, {Moitinho}, {Muinonen},
  {Osinde}, {Pancino}, {Pauwels}, {Petit}, {Recio-Blanco}, {Robin}, {Sarro},
  {Siopis}, {Smith}, {Smith}, {Sozzetti}, {Thuillot}, {van Reeven}, {Viala},
  {Abbas}, {Abreu Aramburu}, {Accart}, {Aguado}, {Allan}, {Allasia},
  {Altavilla}, {{\'A}lvarez}, {Alves}, {Anderson}, {Andrei}, {Anglada Varela},
  {Antiche}, {Antoja}, {Ant{\'o}n}, {Arcay}, {Atzei}, {Ayache}, {Bach},
  {Baker}, {Balaguer-N{\'u}{\~n}ez}, {Barache}, {Barata}, {Barbier}, {Barblan},
  {Baroni}, {Barrado y Navascu{\'e}s}, {Barros}, {Barstow}, {Becciani},
  {Bellazzini}, {Bellei}, {Bello Garc{\'\i}a}, {Belokurov}, {Bendjoya},
  {Berihuete}, {Bianchi}, {Bienaym{\'e}}, {Billebaud}, {Blagorodnova},
  {Blanco-Cuaresma}, {Boch}, {Bombrun}, {Borrachero}, {Bouquillon}, {Bourda},
  {Bouy}, {Bragaglia}, {Breddels}, {Brouillet}, {Br{\"u}semeister},
  {Bucciarelli}, {Budnik}, {Burgess}, {Burgon}, {Burlacu}, {Busonero}, {Buzzi},
  {Caffau}, {Cambras}, {Campbell}, {Cancelliere}, {Cantat-Gaudin}, {Carlucci},
  {Carrasco}, {Castellani}, {Charlot}, {Charnas}, {Charvet}, {Chassat},
  {Chiavassa}, {Clotet}, {Cocozza}, {Collins}, {Collins}, {Costigan}, {Crifo},
  {Cross}, {Crosta}, {Crowley}, {Dafonte}, {Damerdji}, {Dapergolas}, {David},
  {David}, {De Cat}, {de Felice}, {de Laverny}, {De Luise}, {De March}, {de
  Martino}, {de Souza}, {Debosscher}, {del Pozo}, {Delbo}, {Delgado},
  {Delgado}, {di Marco}, {Di Matteo}, {Diakite}, {Distefano}, {Dolding}, {Dos
  Anjos}, {Drazinos}, {Dur{\'a}n}, {Dzigan}, {Ecale}, {Edvardsson}, {Enke},
  {Erdmann}, {Escolar}, {Espina}, {Evans}, {Eynard Bontemps}, {Fabre},
  {Fabrizio}, {Faigler}, {Falc{\~a}o}, {Farr{\`a}s Casas}, {Faye}, {Federici},
  {Fedorets}, {Fern{\'a}ndez-Hern{\'a}ndez}, {Fernique}, {Fienga}, {Figueras},
  {Filippi}, {Findeisen}, {Fonti}, {Fouesneau}, {Fraile}, {Fraser}, {Fuchs},
  {Furnell}, {Gai}, {Galleti}, {Galluccio}, {Garabato}, {Garc{\'\i}a-Sedano},
  {Gar{\'e}}, {Garofalo}, {Garralda}, {Gavras}, {Gerssen}, {Geyer}, {Gilmore},
  {Girona}, {Giuffrida}, {Gomes}, {Gonz{\'a}lez-Marcos},
  {Gonz{\'a}lez-N{\'u}{\~n}ez}, {Gonz{\'a}lez-Vidal}, {Granvik}, {Guerrier},
  {Guillout}, {Guiraud}, {G{\'u}rpide}, {Guti{\'e}rrez-S{\'a}nchez}, {Guy},
  {Haigron}, {Hatzidimitriou}, {Haywood}, {Heiter}, {Helmi}, {Hobbs},
  {Hofmann}, {Holl}, {Holland }, {Hunt}, {Hypki}, {Icardi}, {Irwin}, {Jevardat
  de Fombelle}, {Jofr{\'e}}, {Jonker}, {Jorissen}, {Julbe}, {Karampelas},
  {Kochoska}, {Kohley}, {Kolenberg}, {Kontizas}, {Koposov}, {Kordopatis},
  {Koubsky}, {Kowalczyk}, {Krone-Martins}, {Kudryashova}, {Kull}, {Bachchan},
  {Lacoste-Seris}, {Lanza}, {Lavigne}, {Le Poncin-Lafitte}, {Lebreton},
  {Lebzelter}, {Leccia}, {Leclerc}, {Lecoeur-Taibi}, {Lemaitre}, {Lenhardt},
  {Leroux}, {Liao}, {Licata}, {Lindstr{\o}m}, {Lister}, {Livanou}, {Lobel},
  {L{\"o}ffler}, {L{\'o}pez}, {Lopez-Lozano}, {Lorenz}, {Loureiro},
  {MacDonald}, {Magalh{\~a}es Fernandes}, {Managau}, {Mann}, {Mantelet},
  {Marchal}, {Marchant}, {Marconi}, {Marie}, {Marinoni}, {Marrese},
  {Marschalk{\'o}}, {Marshall}, {Mart{\'\i}n-Fleitas}, {Martino}, {Mary},
  {Matijevi{\v{c}}}, {Mazeh}, {McMillan}, {Messina}, {Mestre}, {Michalik},
  {Millar}, {Miranda}, {Molina}, {Molinaro}, {Molinaro}, {Moln{\'a}r},
  {Moniez}, {Montegriffo}, {Monteiro}, {Mor}, {Mora}, {Morbidelli}, {Morel},
  {Morgenthaler}, {Morley}, {Morris}, {Mulone}, {Muraveva}, {Musella},
  {Narbonne}, {Nelemans}, {Nicastro}, {Noval}, {Ord{\'e}novic},
  {Ordieres-Mer{\'e}}, {Osborne}, {Pagani}, {Pagano}, {Pailler}, {Palacin},
  {Palaversa}, {Parsons}, {Paulsen}, {Pecoraro}, {Pedrosa}, {Pentik{\"a}inen},
  {Pereira}, {Pichon}, {Piersimoni}, {Pineau}, {Plachy}, {Plum}, {Poujoulet},
  {Pr{\v{s}}a}, {Pulone}, {Ragaini}, {Rago}, {Rambaux}, {Ramos-Lerate},
  {Ranalli}, {Rauw}, {Read}, {Regibo}, {Renk}, {Reyl{\'e}}, {Ribeiro},
  {Rimoldini}, {Ripepi}, {Riva}, {Rixon}, {Roelens}, {Romero-G{\'o}mez},
  {Rowell}, {Royer}, {Rudolph}, {Ruiz-Dern}, {Sadowski}, {Sagrist{\`a}
  Sell{\'e}s}, {Sahlmann}, {Salgado}, {Salguero}, {Sarasso}, {Savietto},
  {Schnorhk}, {Schultheis}, {Sciacca}, {Segol}, {Segovia}, {Segransan},
  {Serpell}, {Shih}, {Smareglia}, {Smart}, {Smith}, {Solano}, {Solitro},
  {Sordo}, {Soria Nieto}, {Souchay}, {Spagna}, {Spoto}, {Stampa}, {Steele},
  {Steidelm{\"u}ller}, {Stephenson}, {Stoev}, {Suess}, {S{\"u}veges}, {Surdej},
  {Szabados}, {Szegedi-Elek}, {Tapiador}, {Taris}, {Tauran}, {Taylor},
  {Teixeira}, {Terrett}, {Tingley}, {Trager}, {Turon}, {Ulla}, {Utrilla},
  {Valentini}, {van Elteren}, {Van Hemelryck}, {van Leeuwen}, {Varadi},
  {Vecchiato}, {Veljanoski}, {Via}, {Vicente}, {Vogt}, {Voss}, {Votruba},
  {Voutsinas}, {Walmsley}, {Weiler}, {Weingrill}, {Werner}, {Wevers},
  {Whitehead}, {Wyrzykowski}, {Yoldas}, {{\v{Z}}erjal}, {Zucker}, {Zurbach},
  {Zwitter}, {Alecu}, {Allen}, {Allende Prieto}, {Amorim},
  {Anglada-Escud{\'e}}, {Arsenijevic}, {Azaz}, {Balm}, {Beck}, {Bernstein},
  {Bigot}, {Bijaoui}, {Blasco}, {Bonfigli}, {Bono}, {Boudreault}, {Bressan},
  {Brown}, {Brunet}, {Bunclark}, {Buonanno}, {Butkevich}, {Carret}, {Carrion},
  {Chemin}, {Ch{\'e}reau}, {Corcione}, {Darmigny}, {de Boer}, {de Teodoro}, {de
  Zeeuw}, {Delle Luche}, {Domingues}, {Dubath}, {Fodor}, {Fr{\'e}zouls},
  {Fries}, {Fustes}, {Fyfe}, {Gallardo}, {Gallegos}, {Gardiol}, {Gebran},
  {Gomboc}, {G{\'o}mez}, {Grux}, {Gueguen}, {Heyrovsky}, {Hoar}, {Iannicola},
  {Isasi Parache}, {Janotto}, {Joliet}, {Jonckheere}, {Keil}, {Kim},
  {Klagyivik}, {Klar}, {Knude}, {Kochukhov}, {Kolka}, {Kos}, {Kutka}, {Lainey},
  {LeBouquin}, {Liu}, {Loreggia}, {Makarov}, {Marseille}, {Martayan},
  {Martinez-Rubi}, {Massart}, {Meynadier}, {Mignot}, {Munari}, {Nguyen},
  {Nordlander}, {Ocvirk}, {O'Flaherty}, {Olias Sanz}, {Ortiz}, {Osorio},
  {Oszkiewicz}, {Ouzounis}, {Palmer}, {Park}, {Pasquato}, {Peltzer}, {Peralta},
  {P{\'e}turaud}, {Pieniluoma}, {Pigozzi}, {Poels}, {Prat}, {Prod'homme},
  {Raison}, {Rebordao}, {Risquez}, {Rocca-Volmerange}, {Rosen}, {Ruiz-Fuertes},
  {Russo}, {Sembay}, {Serraller Vizcaino}, {Short}, {Siebert}, {Silva},
  {Sinachopoulos}, {Slezak}, {Soffel}, {Sosnowska}, {Strai{\v{z}}ys}, {ter
  Linden}, {Terrell}, {Theil}, {Tiede}, {Troisi}, {Tsalmantza}, {Tur},
  {Vaccari}, {Vachier}, {Valles}, {Van Hamme}, {Veltz}, {Virtanen}, {Wallut},
  {Wichmann}, {Wilkinson}, {Ziaeepour}, \& {Zschocke}}]{Gaia2016}
{Gaia Collaboration}, {Prusti}, T., {de Bruijne}, J.~H.~J., {et~al.} 2016,
  \aap, 595, A1, \dodoi{10.1051/0004-6361/201629272}

\bibitem[{{Gaia Collaboration} {et~al.}(2018){Gaia Collaboration}, {Brown},
  {Vallenari}, {Prusti}, {de Bruijne}, {Babusiaux}, {Bailer-Jones}, {Biermann},
  {Evans}, {Eyer}, {Jansen}, {Jordi}, {Klioner}, {Lammers}, {Lindegren},
  {Luri}, {Mignard}, {Panem}, {Pourbaix}, {Randich}, {Sartoretti}, {Siddiqui},
  {Soubiran}, {van Leeuwen}, {Walton}, {Arenou}, {Bastian}, {Cropper},
  {Drimmel}, {Katz}, {Lattanzi}, {Bakker}, {Cacciari}, {Casta{\~n}eda},
  {Chaoul}, {Cheek}, {De Angeli}, {Fabricius}, {Guerra}, {Holl}, {Masana},
  {Messineo}, {Mowlavi}, {Nienartowicz}, {Panuzzo}, {Portell}, {Riello},
  {Seabroke}, {Tanga}, {Th{\'e}venin}, {Gracia-Abril}, {Comoretto},
  {Garcia-Reinaldos}, {Teyssier}, {Altmann}, {Andrae}, {Audard},
  {Bellas-Velidis}, {Benson}, {Berthier}, {Blomme}, {Burgess}, {Busso},
  {Carry}, {Cellino}, {Clementini}, {Clotet}, {Creevey}, {Davidson}, {De
  Ridder}, {Delchambre}, {Dell'Oro}, {Ducourant},
  {Fern{\'a}ndez-Hern{\'a}ndez}, {Fouesneau}, {Fr{\'e}mat}, {Galluccio},
  {Garc{\'\i}a-Torres}, {Gonz{\'a}lez-N{\'u}{\~n}ez}, {Gonz{\'a}lez-Vidal},
  {Gosset}, {Guy}, {Halbwachs}, {Hambly}, {Harrison}, {Hern{\'a}ndez},
  {Hestroffer}, {Hodgkin}, {Hutton}, {Jasniewicz}, {Jean-Antoine-Piccolo},
  {Jordan}, {Korn}, {Krone-Martins}, {Lanzafame}, {Lebzelter}, {L{\"o}ffler},
  {Manteiga}, {Marrese}, {Mart{\'\i}n-Fleitas}, {Moitinho}, {Mora}, {Muinonen},
  {Osinde}, {Pancino}, {Pauwels}, {Petit}, {Recio-Blanco}, {Richards},
  {Rimoldini}, {Robin}, {Sarro}, {Siopis}, {Smith}, {Sozzetti}, {S{\"u}veges},
  {Torra}, {van Reeven}, {Abbas}, {Abreu Aramburu}, {Accart}, {Aerts},
  {Altavilla}, {{\'A}lvarez}, {Alvarez}, {Alves}, {Anderson}, {Andrei},
  {Anglada Varela}, {Antiche}, {Antoja}, {Arcay}, {Astraatmadja}, {Bach},
  {Baker}, {Balaguer-N{\'u}{\~n}ez}, {Balm}, {Barache}, {Barata}, {Barbato},
  {Barblan}, {Barklem}, {Barrado}, {Barros}, {Barstow}, {Bartholom{\'e}
  Mu{\~n}oz}, {Bassilana}, {Becciani}, {Bellazzini}, {Berihuete}, {Bertone},
  {Bianchi}, {Bienaym{\'e}}, {Blanco-Cuaresma}, {Boch}, {Boeche}, {Bombrun},
  {Borrachero}, {Bossini}, {Bouquillon}, {Bourda}, {Bragaglia}, {Bramante},
  {Breddels}, {Bressan}, {Brouillet}, {Br{\"u}semeister}, {Brugaletta},
  {Bucciarelli}, {Burlacu}, {Busonero}, {Butkevich}, {Buzzi}, {Caffau},
  {Cancelliere}, {Cannizzaro}, {Cantat-Gaudin}, {Carballo}, {Carlucci},
  {Carrasco}, {Casamiquela}, {Castellani}, {Castro-Ginard}, {Charlot},
  {Chemin}, {Chiavassa}, {Cocozza}, {Costigan}, {Cowell}, {Crifo}, {Crosta},
  {Crowley}, {Cuypers}, {Dafonte}, {Damerdji}, {Dapergolas}, {David}, {David},
  {de Laverny}, {De Luise}, {De March}, {de Martino}, {de Souza}, {de Torres},
  {Debosscher}, {del Pozo}, {Delbo}, {Delgado}, {Delgado}, {Di Matteo},
  {Diakite}, {Diener}, {Distefano}, {Dolding}, {Drazinos}, {Dur{\'a}n},
  {Edvardsson}, {Enke}, {Eriksson}, {Esquej}, {Eynard Bontemps}, {Fabre},
  {Fabrizio}, {Faigler}, {Falc{\~a}o}, {Farr{\`a}s Casas}, {Federici},
  {Fedorets}, {Fernique}, {Figueras}, {Filippi}, {Findeisen}, {Fonti},
  {Fraile}, {Fraser}, {Fr{\'e}zouls}, {Gai}, {Galleti}, {Garabato},
  {Garc{\'\i}a-Sedano}, {Garofalo}, {Garralda}, {Gavel}, {Gavras}, {Gerssen},
  {Geyer}, {Giacobbe}, {Gilmore}, {Girona}, {Giuffrida}, {Glass}, {Gomes},
  {Granvik}, {Gueguen}, {Guerrier}, {Guiraud}, {Guti{\'e}rrez-S{\'a}nchez},
  {Haigron}, {Hatzidimitriou}, {Hauser}, {Haywood}, {Heiter}, {Helmi}, {Heu},
  {Hilger}, {Hobbs}, {Hofmann}, {Holland}, {Huckle}, {Hypki}, {Icardi},
  {Jan{\ss}en}, {Jevardat de Fombelle}, {Jonker}, {Juh{\'a}sz}, {Julbe},
  {Karampelas}, {Kewley}, {Klar}, {Kochoska}, {Kohley}, {Kolenberg},
  {Kontizas}, {Kontizas}, {Koposov}, {Kordopatis}, {Kostrzewa-Rutkowska},
  {Koubsky}, {Lambert}, {Lanza}, {Lasne}, {Lavigne}, {Le Fustec}, {Le
  Poncin-Lafitte}, {Lebreton}, {Leccia}, {Leclerc}, {Lecoeur-Taibi},
  {Lenhardt}, {Leroux}, {Liao}, {Licata}, {Lindstr{\o}m}, {Lister}, {Livanou},
  {Lobel}, {L{\'o}pez}, {Managau}, {Mann}, {Mantelet}, {Marchal}, {Marchant},
  {Marconi}, {Marinoni}, {Marschalk{\'o}}, {Marshall}, {Martino}, {Marton},
  {Mary}, {Massari}, {Matijevi{\v{c}}}, {Mazeh}, {McMillan}, {Messina},
  {Michalik}, {Millar}, {Molina}, {Molinaro}, {Moln{\'a}r}, {Montegriffo},
  {Mor}, {Morbidelli}, {Morel}, {Morris}, {Mulone}, {Muraveva}, {Musella},
  {Nelemans}, {Nicastro}, {Noval}, {O'Mullane}, {Ord{\'e}novic},
  {Ord{\'o}{\~n}ez-Blanco}, {Osborne}, {Pagani}, {Pagano}, {Pailler},
  {Palacin}, {Palaversa}, {Panahi}, {Pawlak}, {Piersimoni}, {Pineau}, {Plachy},
  {Plum}, {Poggio}, {Poujoulet}, {Pr{\v{s}}a}, {Pulone}, {Racero}, {Ragaini},
  {Rambaux}, {Ramos-Lerate}, {Regibo}, {Reyl{\'e}}, {Riclet}, {Ripepi}, {Riva},
  {Rivard}, {Rixon}, {Roegiers}, {Roelens}, {Romero-G{\'o}mez}, {Rowell},
  {Royer}, {Ruiz-Dern}, {Sadowski}, {Sagrist{\`a} Sell{\'e}s}, {Sahlmann},
  {Salgado}, {Salguero}, {Sanna}, {Santana-Ros}, {Sarasso}, {Savietto},
  {Schultheis}, {Sciacca}, {Segol}, {Segovia}, {S{\'e}gransan}, {Shih},
  {Siltala}, {Silva}, {Smart}, {Smith}, {Solano}, {Solitro}, {Sordo}, {Soria
  Nieto}, {Souchay}, {Spagna}, {Spoto}, {Stampa}, {Steele},
  {Steidelm{\"u}ller}, {Stephenson}, {Stoev}, {Suess}, {Surdej}, {Szabados},
  {Szegedi-Elek}, {Tapiador}, {Taris}, {Tauran}, {Taylor}, {Teixeira},
  {Terrett}, {Teyssand ier}, {Thuillot}, {Titarenko}, {Torra Clotet}, {Turon},
  {Ulla}, {Utrilla}, {Uzzi}, {Vaillant}, {Valentini}, {Valette}, {van Elteren},
  {Van Hemelryck}, {van Leeuwen}, {Vaschetto}, {Vecchiato}, {Veljanoski},
  {Viala}, {Vicente}, {Vogt}, {von Essen}, {Voss}, {Votruba}, {Voutsinas},
  {Walmsley}, {Weiler}, {Wertz}, {Wevers}, {Wyrzykowski}, {Yoldas},
  {{\v{Z}}erjal}, {Ziaeepour}, {Zorec}, {Zschocke}, {Zucker}, {Zurbach}, \&
  {Zwitter}}]{Gaia2018}
{Gaia Collaboration}, {Brown}, A.~G.~A., {Vallenari}, A., {et~al.} 2018, \aap,
  616, A1, \dodoi{10.1051/0004-6361/201833051}

\bibitem[{{Gianotto} {et~al.}(2025){Gianotto}, {Carbognani}, {Fenucci},
  {Devog{\`e}le}, {Ramirez-Moreta}, {Micheli}, {Salerno}, {Santana-Ros},
  {Cano}, {Conversi}, {Drury}, {Faggioli}, {F{\"o}hring}, {Kresken},
  {Machnitzky}, {Moissl}, {Oca{\~n}a}, {Oliviero}, {Alonso-Peleato},
  {Revellino}, \& {Rudawska}}]{Gianotto2025}
{Gianotto}, F., {Carbognani}, A., {Fenucci}, M., {et~al.} 2025, arXiv e-prints,
  arXiv:2502.09712, \dodoi{10.48550/arXiv.2502.09712}

\bibitem[{{Granvik} {et~al.}(2016){Granvik}, {Morbidelli}, {Jedicke}, {Bolin},
  {Bottke}, {Beshore}, {Vokrouhlick{\'y}}, {Delb{\`o}}, \&
  {Michel}}]{Granvik2016}
{Granvik}, M., {Morbidelli}, A., {Jedicke}, R., {et~al.} 2016, \nat, 530, 303,
  \dodoi{10.1038/nature16934}

\bibitem[{{Hanu{\v{s}}} {et~al.}(2017){Hanu{\v{s}}}, {Viikinkoski}, {Marchis},
  {{\v{D}}urech}, {Kaasalainen}, {Delbo'}, {Herald}, {Frappa}, {Hayamizu},
  {Kerr}, {Preston}, {Timerson}, {Dunham}, \& {Talbot}}]{Hanus2017density}
{Hanu{\v{s}}}, J., {Viikinkoski}, M., {Marchis}, F., {et~al.} 2017, \aap, 601,
  A114, \dodoi{10.1051/0004-6361/201629956}

\bibitem[{{Harris} \& {Lagerros}(2002)}]{Harris2002}
{Harris}, A.~W., \& {Lagerros}, J.~S.~V. 2002, Asteroids III, 205

\bibitem[{{Hudson} \& {Ostro}(1999)}]{Hudson1999}
{Hudson}, R.~S., \& {Ostro}, S.~J. 1999, \icarus, 140, 369,
  \dodoi{10.1006/icar.1999.6142}

\bibitem[{{Huehnerhoff} {et~al.}(2016){Huehnerhoff}, {Ketzeback}, {Bradley},
  {Dembicky}, {Doughty}, {Hawley}, {Johnson}, {Klaene}, {Leon}, {McMillan},
  {Owen}, {Sayres}, {Sheen}, \& {Shugart}}]{Huehnerhoff2016}
{Huehnerhoff}, J., {Ketzeback}, W., {Bradley}, A., {et~al.} 2016, in \procspie,
  Vol. 9908, Ground-based and Airborne Instrumentation for Astronomy VI, 99085H

\bibitem[{{Ivezi{\'c}} {et~al.}(2001){Ivezi{\'c}}, {Tabachnik}, {Rafikov},
  {Lupton}, {Quinn}, {Hammergren}, {Eyer}, {Chu}, {Armstrong}, {Fan},
  {Finlator}, {Geballe}, {Gunn}, {Hennessy}, {Knapp}, {Leggett}, {Munn},
  {Pier}, {Rockosi}, {Schneider}, {Strauss}, {Yanny}, {Brinkmann}, {Csabai},
  {Hindsley}, {Kent}, {Lamb}, {Margon}, {McKay}, {Smith}, {Waddel}, {York}, \&
  {SDSS Collaboration}}]{Ivezic2001}
{Ivezi{\'c}}, {\v Z}., {Tabachnik}, S., {Rafikov}, R., {et~al.} 2001, \aj, 122,
  2749, \dodoi{10.1086/323452}

\bibitem[{{Ivezi{\'c}} {et~al.}(2002){Ivezi{\'c}}, {Lupton}, {Juri{\'c}},
  {Tabachnik}, {Quinn}, {Gunn}, {Knapp}, {Rockosi}, \&
  {Brinkmann}}]{Ivezic2002}
{Ivezi{\'c}}, {\v Z}., {Lupton}, R.~H., {Juri{\'c}}, M., {et~al.} 2002, \aj,
  124, 2943, \dodoi{10.1086/344077}

\bibitem[{{Jenniskens} {et~al.}(2009){Jenniskens}, {Shaddad}, {Numan}, {Elsir},
  {Kudoda}, {Zolensky}, {Le}, {Robinson}, {Friedrich}, {Rumble}, {Steele},
  {Chesley}, {Fitzsimmons}, {Duddy}, {Hsieh}, {Ramsay}, {Brown}, {Edwards},
  {Tagliaferri}, {Boslough}, {Spalding}, {Dantowitz}, {Kozubal}, {Pravec},
  {Borovicka}, {Charvat}, {Vaubaillon}, {Kuiper}, {Albers}, {Bishop},
  {Mancinelli}, {Sandford}, {Milam}, {Nuevo}, \& {Worden}}]{Jenniskens2009}
{Jenniskens}, P., {Shaddad}, M.~H., {Numan}, D., {et~al.} 2009, \nat, 458, 485,
  \dodoi{10.1038/nature07920}

\bibitem[{{Jenniskens} {et~al.}(2021){Jenniskens}, {Gabadirwe}, {Yin},
  {Proyer}, {Moses}, {Kohout}, {Franchi}, {Gibson}, {Kowalski}, {Christensen},
  {Gibbs}, {Heinze}, {Denneau}, {Farnocchia}, {Chodas}, {Gray}, {Micheli},
  {Moskovitz}, {Onken}, {Wolf}, {Devillepoix}, {Ye}, {Robertson}, {Brown},
  {Lyytinen}, {Moilanen}, {Albers}, {Cooper}, {Assink}, {Evers}, {Lahtinen},
  {Seitshiro}, {Laubenstein}, {Wantlo}, {Moleje}, {Maritinkole}, {Suhonen},
  {Zolensky}, {Ashwal}, {Hiroi}, {Sears}, {Sehlke}, {Maturilli}, {Sanborn},
  {Huyskens}, {Dey}, {Ziegler}, {Busemann}, {Riebe}, {Meier}, {Welten},
  {Caffee}, {Zhou}, {Li}, {Li}, {Liu}, {Tang}, {McLain}, {Dworkin}, {Glavin},
  {Schmitt-Kopplin}, {Sabbah}, {Joblin}, {Granvik}, {Mosarwa}, \&
  {Botepe}}]{Jenniskens2021}
{Jenniskens}, P., {Gabadirwe}, M., {Yin}, Q.-Z., {et~al.} 2021, \maps, 56, 844,
  \dodoi{10.1111/maps.13653}

\bibitem[{Jester {et~al.}(2005)Jester, Schneider, Richards, Green, Schmidt,
  Hall, Strauss, Vanden~Berk, Stoughton, Gunn, Brinkmann, Kent, Smith, Tucker,
  \& Yanny}]{Jester_2005}
Jester, S., Schneider, D.~P., Richards, G.~T., {et~al.} 2005, The Astronomical
  Journal, 130, 873, \dodoi{10.1086/432466}

\bibitem[{{Juri{\'c}} {et~al.}(2002){Juri{\'c}}, {Ivezi{\'c}}, {Lupton},
  {Quinn}, {Tabachnik}, {Fan}, {Gunn}, {Hennessy}, {Knapp}, {Munn}, {Pier},
  {Rockosi}, {Schneider}, {Brinkmann}, {Csabai}, \& {Fukugita}}]{Juric2002}
{Juri{\'c}}, M., {Ivezi{\'c}}, {\v Z}., {Lupton}, R.~H., {et~al.} 2002, \aj,
  124, 1776, \dodoi{10.1086/341950}

\bibitem[{{Kareta} {et~al.}(2024){Kareta}, {Vida}, {Micheli}, {Moskovitz},
  {Wiegert}, {Brown}, {McCausland}, {Devillepoix}, {Male{\v{c}}i{\'c}},
  {Prtenjak}, {{\v{S}}egon}, {Shafransky}, \& {Farnocchia}}]{Kareta2024}
{Kareta}, T., {Vida}, D., {Micheli}, M., {et~al.} 2024, \psj, 5, 253,
  \dodoi{10.3847/PSJ/ad8b22}

\bibitem[{{Larson} {et~al.}(1998){Larson}, {Brownlee}, {Hergenrother}, \&
  {Spahr}}]{Larson1998}
{Larson}, S., {Brownlee}, J., {Hergenrother}, C., \& {Spahr}, T. 1998, in
  Bulletin of the American Astronomical Society, Vol.~30, 1037

\bibitem[{{Lomb}(1976)}]{Lomb1976}
{Lomb}, N.~R. 1976, \apss, 39, 447, \dodoi{10.1007/BF00648343}

\bibitem[{{Mann} {et~al.}(2007){Mann}, {Jewitt}, \& {Lacerda}}]{Mann2007}
{Mann}, R.~K., {Jewitt}, D., \& {Lacerda}, P. 2007, \aj, 134, 1133,
  \dodoi{10.1086/520328}

\bibitem[{{Masiero} {et~al.}(2024){Masiero}, {Kwon}, {Dahlen}, {Masci}, \&
  {Mainzer}}]{Masiero2024}
{Masiero}, J.~R., {Kwon}, Y.~G., {Dahlen}, D.~W., {Masci}, F.~J., \& {Mainzer},
  A.~K. 2024, \psj, 5, 113, \dodoi{10.3847/PSJ/ad42a2}

\bibitem[{{Nesvorn{\'y}} {et~al.}(2023){Nesvorn{\'y}}, {Deienno}, {Bottke},
  {Jedicke}, {Naidu}, {Chesley}, {Chodas}, {Granvik}, {Vokrouhlick{\'y}},
  {Bro{\v{z}}}, {Morbidelli}, {Christensen}, {Shelly}, \&
  {Bolin}}]{Nesvorny2023NEOMOD}
{Nesvorn{\'y}}, D., {Deienno}, R., {Bottke}, W.~F., {et~al.} 2023, \aj, 166,
  55, \dodoi{10.3847/1538-3881/ace040}

\bibitem[{{Nesvorn{\'y}} {et~al.}(2024){Nesvorn{\'y}}, {Vokrouhlick{\'y}},
  {Shelly}, {Deienno}, {Bottke}, {Fuls}, {Jedicke}, {Naidu}, {Chesley},
  {Chodas}, {Farnocchia}, \& {Delbo}}]{Nesvorny2024NEOMOD3}
{Nesvorn{\'y}}, D., {Vokrouhlick{\'y}}, D., {Shelly}, F., {et~al.} 2024,
  \icarus, 417, 116110, \dodoi{10.1016/j.icarus.2024.116110}

\bibitem[{{Pentland} {et~al.}(2006){Pentland}, {Gonzales}, {Harris}, E.V., \&
  {Downey}}]{Pentland2006}
{Pentland}, G., {Gonzales}, K., {Harris}, K., E.V., R., \& {Downey}, E. 2006,
  in Ground-based and Airborne Telescopes, ed. L.~M. Stepp, Vol. 6267,
  International Society for Optics and Photonics (SPIE), 62670C.
\newblock \url{https://doi.org/10.1117/12.669795}

\bibitem[{{Purdum} {et~al.}(2021){Purdum}, {Lin}, {Bolin}, {Sharma}, {Choi},
  {Bhalerao}, {Hanu{\v{s}}}, {Kumar}, {Quimby}, {van Roestel}, {Zhai},
  {Fernandez}, {Lisse}, {Bodewits}, {Fremling}, {Ryan Golovich}, {Hsu}, {Ip},
  {Ngeow}, {Saini}, {Shao}, {Yao}, {Ahumada}, {Anand}, {Andreoni}, {Burdge},
  {Burruss}, {Chang}, {Copperwheat}, {Coughlin}, {De}, {Dekany}, {Delacroix},
  {Drake}, {Duev}, {Graham}, {Hale}, {Kool}, {Kasliwal}, {Kostadinova},
  {Kulkarni}, {Laher}, {Mahabal}, {Masci}, {Mr{\'o}z}, {Neill}, {Riddle},
  {Rodriguez}, {Smith}, {Walters}, {Yan}, \& {Zolkower}}]{Purdum2021}
{Purdum}, J.~N., {Lin}, Z.-Y., {Bolin}, B.~T., {et~al.} 2021, \apjl, 911, L35,
  \dodoi{10.3847/2041-8213/abf2ca}

\bibitem[{{Raab}(2012)}]{Raab2012a}
{Raab}, H. 2012, {Astrometrica: Astrometric data reduction of CCD images}.
\newblock \doeprint{1203.012}

\bibitem[{{Spurn{\'y}} {et~al.}(2024){Spurn{\'y}}, {Borovi{\v{c}}ka},
  {Shrben{\'y}}, {Hankey}, \& {Neubert}}]{Spurny2024}
{Spurn{\'y}}, P., {Borovi{\v{c}}ka}, J., {Shrben{\'y}}, L., {Hankey}, M., \&
  {Neubert}, R. 2024, \aap, 686, A67, \dodoi{10.1051/0004-6361/202449735}

\bibitem[{{Stellingwerf}(1978)}]{Stellingwerf1978}
{Stellingwerf}, R.~F. 1978, \apj, 224, 953, \dodoi{10.1086/156444}

\bibitem[{{Tonry} {et~al.}(2012){Tonry}, {Stubbs}, {Lykke}, {Doherty},
  {Shivvers}, {Burgett}, {Chambers}, {Hodapp}, {Kaiser}, {Kudritzki},
  {Magnier}, {Morgan}, {Price}, \& {Wainscoat}}]{Tonry2012}
{Tonry}, J.~L., {Stubbs}, C.~W., {Lykke}, K.~R., {et~al.} 2012, \apj, 750, 99,
  \dodoi{10.1088/0004-637X/750/2/99}

\bibitem[{{Tonry} {et~al.}(2018){Tonry}, {Denneau}, {Heinze}, {Stalder},
  {Smith}, {Smartt}, {Stubbs}, {Weiland}, \& {Rest}}]{Tonry2018}
{Tonry}, J.~L., {Denneau}, L., {Heinze}, A.~N., {et~al.} 2018, \pasp, 130,
  064505, \dodoi{10.1088/1538-3873/aabadf}

\bibitem[{{Vera C. Rubin Observatory LSST Solar System Science Collaboration}
  {et~al.}(2020){Vera C. Rubin Observatory LSST Solar System Science
  Collaboration}, {Jones}, {Bannister}, {Bolin}, {Chandler}, {Chesley}, {Eggl},
  {Greenstreet}, {Holt}, {Hsieh}, {Ivezi{\'c}}, {Juri{\'c}}, {Kelley},
  {Knight}, {Malhotra}, {Oldroyd}, {Sarid}, {Schwamb}, {Snodgrass}, {Solontoi},
  \& {Trilling}}]{Vera2020}
{Vera C. Rubin Observatory LSST Solar System Science Collaboration}, {Jones},
  R.~L., {Bannister}, M.~T., {et~al.} 2020, arXiv e-prints, arXiv:2009.07653,
  \dodoi{10.48550/arXiv.2009.07653}

\bibitem[{{Veres} \& {Green}(2024)}]{VeresCBET2025}
{Veres}, P., \& {Green}, D.~W.~E. 2024, Central Bureau Electronic Telegrams,
  5438

\bibitem[{{Vere{\v s}} {et~al.}(2015){Vere{\v s}}, {Jedicke}, {Fitzsimmons},
  {Denneau}, {Granvik}, {Bolin}, {Chastel}, {Wainscoat}, {Burgett}, {Chambers},
  {Flewelling}, {Kaiser}, {Magnier}, {Morgan}, {Price}, {Tonry}, \&
  {Waters}}]{Veres2015}
{Vere{\v s}}, P., {Jedicke}, R., {Fitzsimmons}, A., {et~al.} 2015, \icarus,
  261, 34, \dodoi{10.1016/j.icarus.2015.08.007}

\bibitem[{{Walsh} {et~al.}(2013){Walsh}, {Delbo}, {Bottke}, {Vokrouhlick{\'y}},
  \& {Lauretta}}]{Walsh2013}
{Walsh}, K.~J., {Delbo}, M., {Bottke}, W.~F., {Vokrouhlick{\'y}}, D., \&
  {Lauretta}, D.~S. 2013, \icarus, 225, 283,
  \dodoi{10.1016/j.icarus.2013.03.005}

\bibitem[{{Whidden} {et~al.}(2019){Whidden}, {Bryce Kalmbach}, {Connolly},
  {Jones}, {Smotherman}, {Bektesevic}, {Slater}, {Becker}, {Ivezi{\'c}},
  {Juri{\'c}}, {Bolin}, {Moeyens}, {F{\"o}rster}, \& {Golkhou}}]{Whidden2019}
{Whidden}, P.~J., {Bryce Kalmbach}, J., {Connolly}, A.~J., {et~al.} 2019, \aj,
  157, 119, \dodoi{10.3847/1538-3881/aafd2d}

\bibitem[{{Wierzchos} {et~al.}(2024){Wierzchos}, {Seaman}, {Fay}, {Grauer},
  {Hogan}, {Larson}, {Rankin}, {Carvajal}, {Shelly}, {Fuls}, {Beuden}, {Gibbs},
  {Fazekas}, {Leonard}, {Groeller}, {Kowalski}, {Ryan}, {Ryan}, {Cromer},
  {Valentine}, {Lessig}, {Goodin}, {Hug}, {Pittichova}, {Santana-Ros},
  {Oca{\~n}a}, {Conversi}, {Micheli}, {Linder}, {Holmes}, {Sato}, {Chambers},
  {Herman}, {Smith}, {Fairlamb}, {Gao}, {Weryk}, {Minguez}, {Schultz}, {Huber},
  {Ramanjooloo}, {Lowe}, {Wainscoat}, {de Boer}, {Magnier}, {Lin}, {Devogele},
  {Fohring}, {Conversi}, {Oca{\~n}a}, {Micheli}, \&
  {Santana-Ros}}]{Wierzchos2024RW1}
{Wierzchos}, K.~W., {Seaman}, R.~L., {Fay}, D., {et~al.} 2024, Minor Planet
  Electronic Circulars, 2024-R68, \dodoi{10.48377/MPEC/2024-R68}

\bibitem[{{Williams}(2024)}]{Williams2024RW1}
{Williams}, G.~V. 2024, Minor Planet Electronic Circulars, 2024-R79,
  \dodoi{10.48377/mpec/2024-r79}

\bibitem[{{Willmer}(2018)}]{Willmer2018}
{Willmer}, C. N.~A. 2018, \apjs, 236, 47, \dodoi{10.3847/1538-4365/aabfdf}

\end{thebibliography}

\clearpage
\newpage
\begin{longtable}{|c|c|c|}
\caption{Summary of \rw photometry from 2025 September 4 UTC.\label{t.photometry1}}\\
\hline
Date$^1$ & Mag$^2$ & Mag unc.$^3$\\
(MJD UTC)&&(s)\\
\hline
\endfirsthead
\multicolumn{3}{c}%
{\tablename\ \thetable\ -- \textit{Continued from previous page}} \\
\hline
Date$^1$ & Mag$^2$ & Mag unc.$^3$ \\
(MJD UTC)&&(s)\\
\hline
\endhead
\endfoot
\hline
\endlastfoot
60557.41511548 & 19.36 & 0.11 \\
60557.41593433 & 19.41 & 0.10 \\
60557.41654590 & 19.60 & 0.09 \\
60557.41783633 & 19.78 & 0.09 \\
60557.42766710 & 19.78 & 0.08 \\
60557.42897012 & 19.93 & 0.08 \\
60557.42971581 & 20.04 & 0.13 \\
60557.43117126 & 20.05 & 0.12 \\
60557.43183214 & 19.90 & 0.07 \\
60557.43287760 & 19.71 & 0.04 \\
60557.43491467 & 19.30 & 0.06 \\
60557.43715681 & 19.41 & 0.06 \\
60557.44587752 & 19.44 & 0.06 \\
60557.44897267 & 19.47 & 0.04 \\
60557.45206661 & 19.79 & 0.05 \\
60557.45515838 & 19.77 & 0.05 \\
\hline
\caption{Columns: (1) observation date; (2) r-band equivalent magnitude; (3) 1-$\sigma$ magnitude uncertainty}
\label{t:photo}
\end{longtable}

\end{document}